\documentclass[aps,reprint,twocolumn,superscriptaddress,citeautoscript,showpacs,amsmath,amssymb,floatfix,prb]{revtex4-1}

\usepackage[T1]{fontenc}
\usepackage[utf8]{inputenc}
\usepackage{color}
\usepackage[dvipsnames]{xcolor}
\usepackage{graphicx}
\usepackage{amsmath}
\usepackage{amssymb}
\usepackage{amsbsy}
\usepackage{amstext}
\usepackage{multirow}
\usepackage{placeins}
\usepackage{makecell}
\usepackage{gensymb}
\usepackage{verbatim}
\usepackage{booktabs}
\usepackage{textcomp}

\usepackage[unicode=true,pdfusetitle, bookmarks=true,bookmarksnumbered=false,bookmarksopen=false, breaklinks=false,pdfborder={0 0 0},colorlinks=true]{hyperref}
\hypersetup{urlcolor=blue, citecolor=blue}

\makeatletter
\newcommand{\lyxmathsym}[1]{\ifmmode\begingroup\def\b@ld{bold}
  \text{\ifx\math@version\b@ld\bfseries\fi#1}\endgroup\else#1\fi}

\makeatother

\begin{document}
\title{Physical properties of ${R_2}${Co$_6$}{Al$_{20-\delta}$} ($R$ = Gd-Tm, Y) single
crystals}
\author{Sushma Kumari}
\affiliation{Department of Physics and Astronomy, Iowa State University, Ames, IA 50011, USA.}
\affiliation{Ames National Laboratory, Iowa State University, Ames, IA 50011, USA.}
\author{Fernando A. Garcia}
\affiliation{Ames National Laboratory, Iowa State University, Ames, IA 50011, USA.}
\affiliation{ Instituto de Física, Universidade de São Paulo, São Paulo-SP, 05508-090, Brazil.}
\author{Juan Schmidt}
\affiliation{Department of Physics and Astronomy, Iowa State University, Ames, IA 50011, USA.}
\affiliation{Ames National Laboratory, Iowa State University, Ames, IA 50011, USA.}
\author{Tyler J. Slade}
\affiliation{Ames National Laboratory, Iowa State University, Ames, IA 50011, USA.}
\author{ Aashish Sapkota}
\affiliation{Ames National Laboratory, Iowa State University, Ames, IA 50011, USA.}
\author{ Ajay Kumar}
\affiliation{Ames National Laboratory, Iowa State University, Ames, IA 50011, USA.}
\author{ Yaroslav Mudryk}
\affiliation{Ames National Laboratory, Iowa State University, Ames, IA 50011, USA.}
\author{Paul C. Canfield}
\affiliation{Department of Physics and Astronomy, Iowa State University, Ames, IA 50011, USA.}
\affiliation{Ames National Laboratory, Iowa State University, Ames, IA 50011, USA.}
\author{Raquel A. Ribeiro}
\affiliation{Department of Physics and Astronomy, Iowa State University, Ames, IA 50011, USA.}
\affiliation{Ames National Laboratory, Iowa State University, Ames, IA 50011, USA.}

\begin{abstract}
Rare-earth ($R$) based intermetallic compounds can often exhibit diverse physical properties and distinct magnetic anisotropies. A Notable example are the light rare earth members of the monoclinic, $R_{2}$Co$_{6}$Al$_{19}$  series that are known to display a range of physical properties, from non-Fermi liquid behavior to antiferromagnetic (AFM) ordering, with properties that vary depending on $R$. In this work, we have extended this series to the heavy rare earths and systematically investigate the synthesis, crystal structure, and physical properties of single crystals of $R_{2}$Co$_{6}$Al$_{20-\delta}$ for $R=$ Gd - Tm and Y. Single crystal X-ray diffraction reveals that these materials adopt an orthorhombic $Imma$-type structure with $\delta$ varying non-monotonically across the heavy rare earths; ranging from 0.73 for Dy to 0.91 for Gd. Temperature-dependent specific heat, resistivity, and magnetization measurements demonstrate AFM ordering in all materials, with the Neel temperature ($T_{\text{N}}$) ranging from $1.8$ K for Ho to $11.8$ K for Tb. Notably, Gd and Tb-based materials exhibit two distinct AFM transitions, separated by approximately $2-3$ K. These findings establish the heavy rare-earth members of the $R_{2}$Co$_{6}$Al$_{20-\delta}$ series as anisotropic antiferromagnets with strong crystal electric field effects and exchange anisotropy. The observed deviation from de Gennes scaling and the anisotropy crossover across the series highlight the important interplay between RKKY exchange and crystal electric field interactions in this orthorhombic system.

\end{abstract}
\maketitle

\section{Introduction}
Magnetic materials with complex structures and distinct magnetic anisotropies often exhibit intriguing physical properties, making them the focus of extensive research. Among these, rare-earth ($R$) based intermetallic compounds have garnered attention because of their unique magnetic behaviors. Lower symmetry structures are of particular interest since they may introduce magnetic anisotropies that are highly desirable for technological applications \citep{lewis_perspectives_2013,mccallum_practical_2014}. 

For $R$-based materials that do not support transition metal based magnetism, the magnetic anisotropy is generally determined by a single-ion effect set by the crystalline electric field (CEF). This is well illustrated by systematic investigations of material series whose magnetism is of purely local moment ($R=$ Pr, Nd and Gd-Tm) character such as $R$Ni$_{2}$Ge$_{2}$ \citep{budko_anisotropy_1999}, $R$Ni$_{2}$B$_{2}$C
\citep{bhatnagar_electrical_1997,canfield_rni2b2c_1997,fisher_anisotropic_1997}, $R$AgSb$_{2}$, \citep{petrovic_anisotropic_2003,myers_systematic_1999}
$R$Ni$_{3}$Ga$_{9}$, \citep{silva_crystal_2017,mendonca_y-substitution_2023} $R$AgGe,\citep{MOROSAN2004298} and $R$Co$_2$Al$_8$\citep{xcvn-4drj} . Whereas CEF effects are often treated phenomenologically, a first-principles theory of CEF effects \citep{lee_toward_2024} is desirable and may impact the design of rare-earth-based materials used for applications.

It was previously recognized that the $R$-Co-Al ternary phase diagram \citep{R2Co6Al9poly} hosts many ternary phases, and the physical properties of many of these phases remain largely unexplored. For example, the $R_{2}\text{Co}_{6}\text{Al}_{19}$ materials were reported for $R=$ Y, Ce, Pr, Sm and Gd through Tm but the limited available data suggest a good potential for further explorations. In fact,  the known phenomenology ranges from non-Fermi liquid in Ce$_{2}$$\text{Co}_{6}\text{Al}_{19}$ to antiferromagnetic (AFM) order in Gd$_{2}$$\text{Co}_{6}\text{Al}_{19}$ \citep{LaPr,Ce}. In addition, the investigation of the $R_{2}\text{Co}_{6}\text{Al}_{19}$ physical properties was limited to polycrystalline samples \citep{R2Co6Al9poly}. 

Motivated by this knowledge gap,  we decided to investigate the growth of $R_{2}\text{Co}_{6}\text{Al}_{19}$ single crystals and here we present a systematic characterization of their crystal structure and physical properties.  All materials were grown out of an Al-rich solution similar to what was recently adopted for obtaining $R\text{Co}_{2}\text{Al}_{8}$ single crystals \citep{treadwell_investigation_2015,watkins-curry_strategic_2015}. However, contrary to prior work on polycrystalline samples and powder X-ray diffraction, which suggested monoclinic $C2/m$ symmetry, we provide X-ray diffraction data that show that these materials adopt an orthorhombic $Imma$ structure with the chemical formula $R_{2}\text{Co}_{6}\text{Al}_{20-\delta}$ ($\delta$ $\approx$ 1). Based upon temperature-dependent specific heat ($C_{p}$), resistivity ($\rho$) and magnetization ($M$) measurements, we show that all $R=$ Gd-Tm materials manifest AFM order with a  Neel transition temperature ($T_{\text{N}}$) in the range of $1.8$ K ($R=$ Ho) to $11.8$ K ($R=$Tb). In addition, the Gd and Tb-based materials show two consecutive AFM transitions, separated by about $2-3$ K.

\section{Experimental techniques}
\begin{figure}
\centering{}\includegraphics[width=0.47\textwidth]{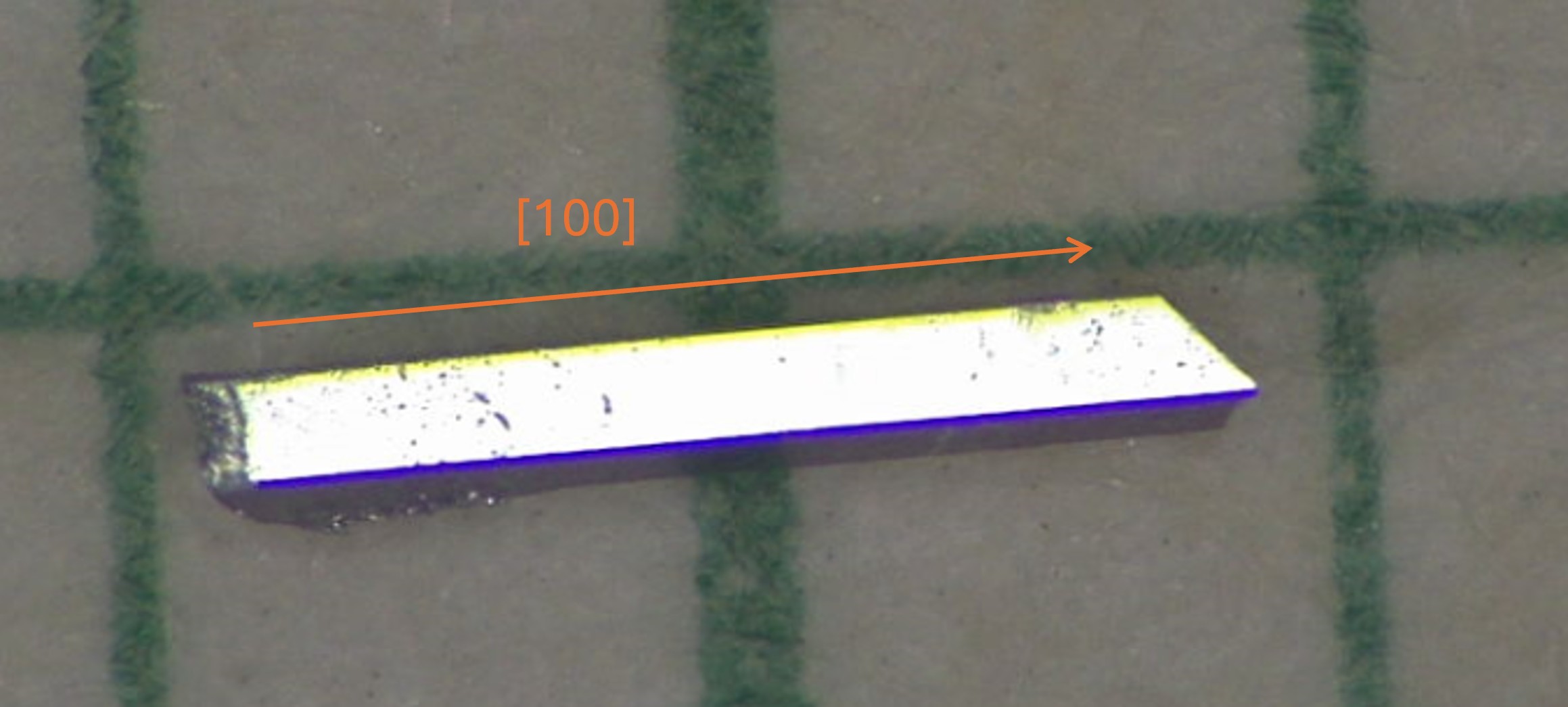} 

\caption{\footnotesize{   Gd$_{2}\text{Co}_{6}\text{Al}_{20-\delta}$ single crystals, photographed over a mm-grid, obtained using the flux method described in the text. We observed that the crystals typically grow as a rod-like shape, where long axis is along the orthorhombic $a$-axis. 
\protect\label{fig:crystal_image}}}
\end{figure}

\begin{figure}
\centering \includegraphics[scale=0.42]{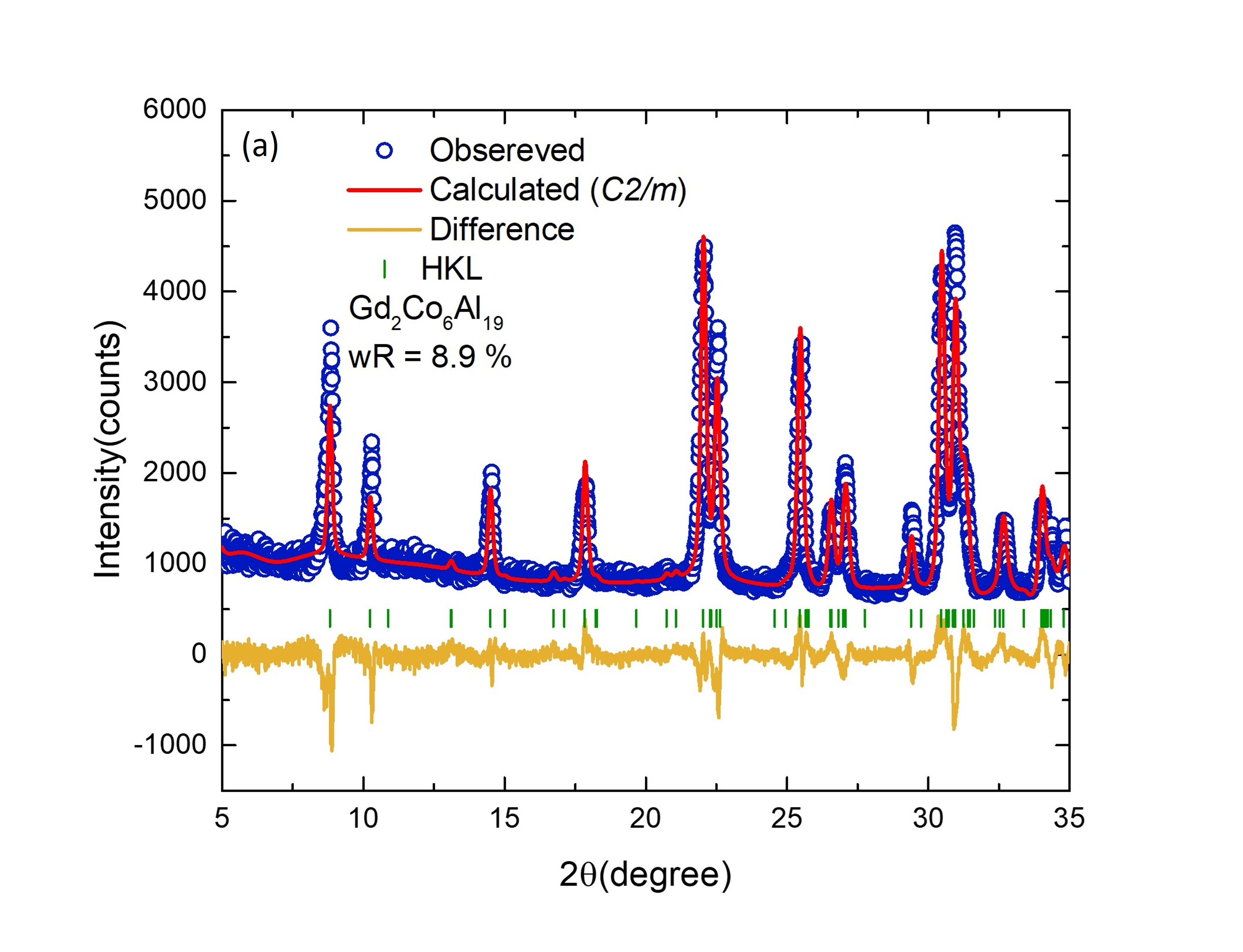}
\centering \includegraphics[scale=0.42]{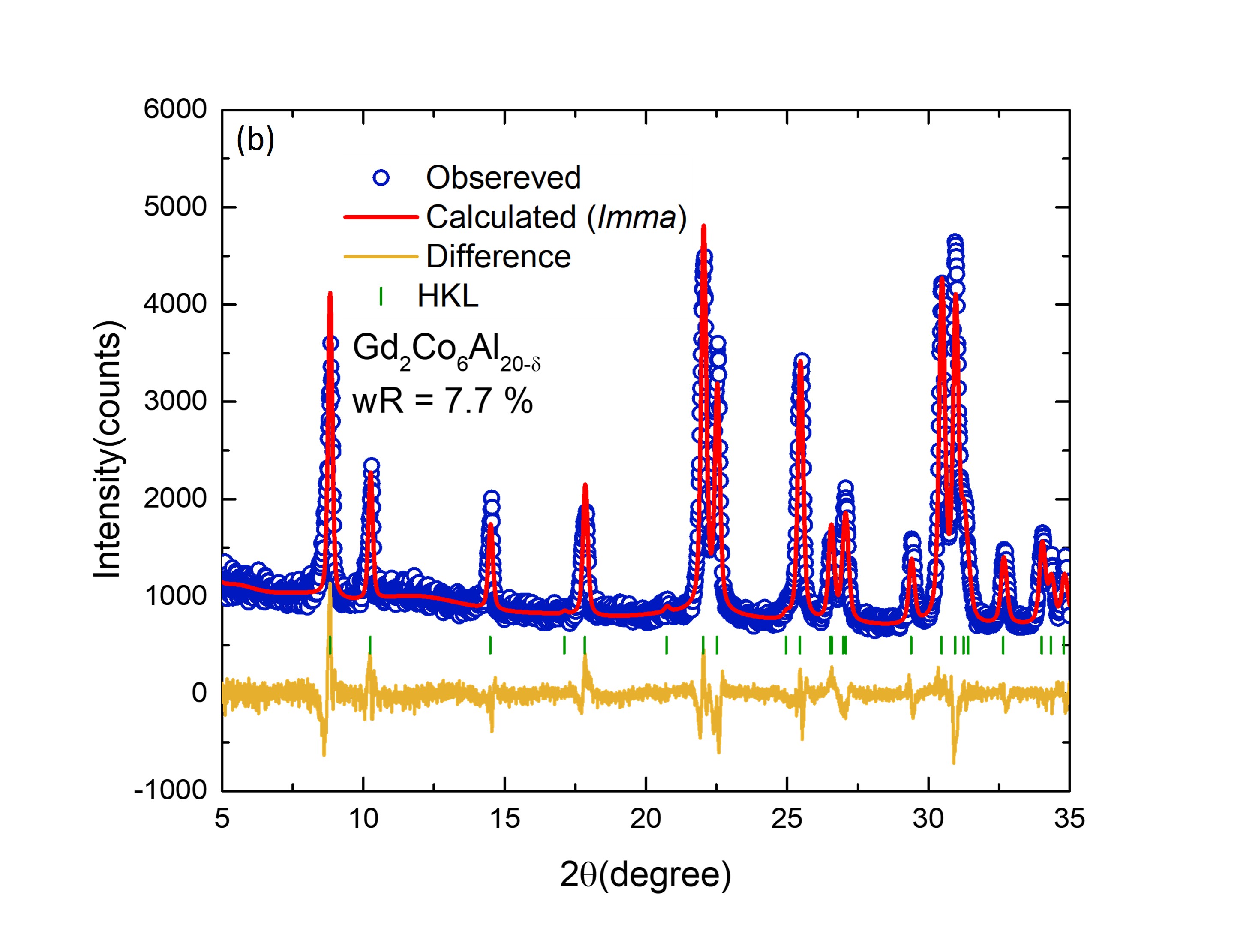}
\includegraphics[scale=0.42]{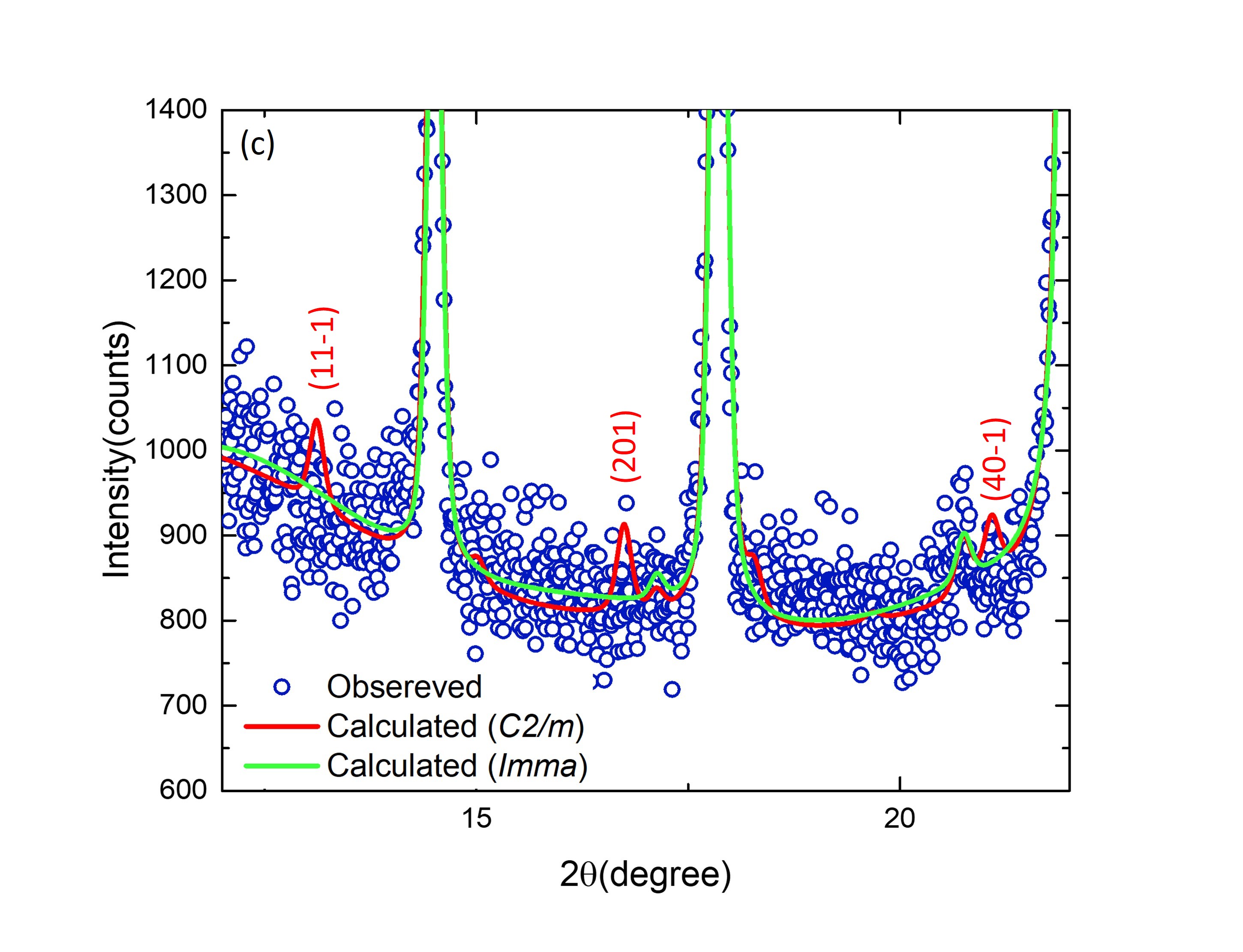}

\caption{\footnotesize{ $a)$ PXRD data of Gd$_2$Co$_6$Al$_{19}$} from 5-35 $\degree$} fitted with space group $C2/m$. $b)$ PXRD data of Gd$_2$Co$_6$Al$_{20-\delta}$ from 5-35 $\degree$ fitted with space group $Imma$. $c)$ Close view of PXRD data fitted with both space groups ($C2/m$ $\&$ $Imma$) from 12-22\degree.
\protect\label{fig:PXRD data of GdCoAl}
\end{figure}

\begin{figure}
\centering{}\includegraphics[width=0.45\textwidth]{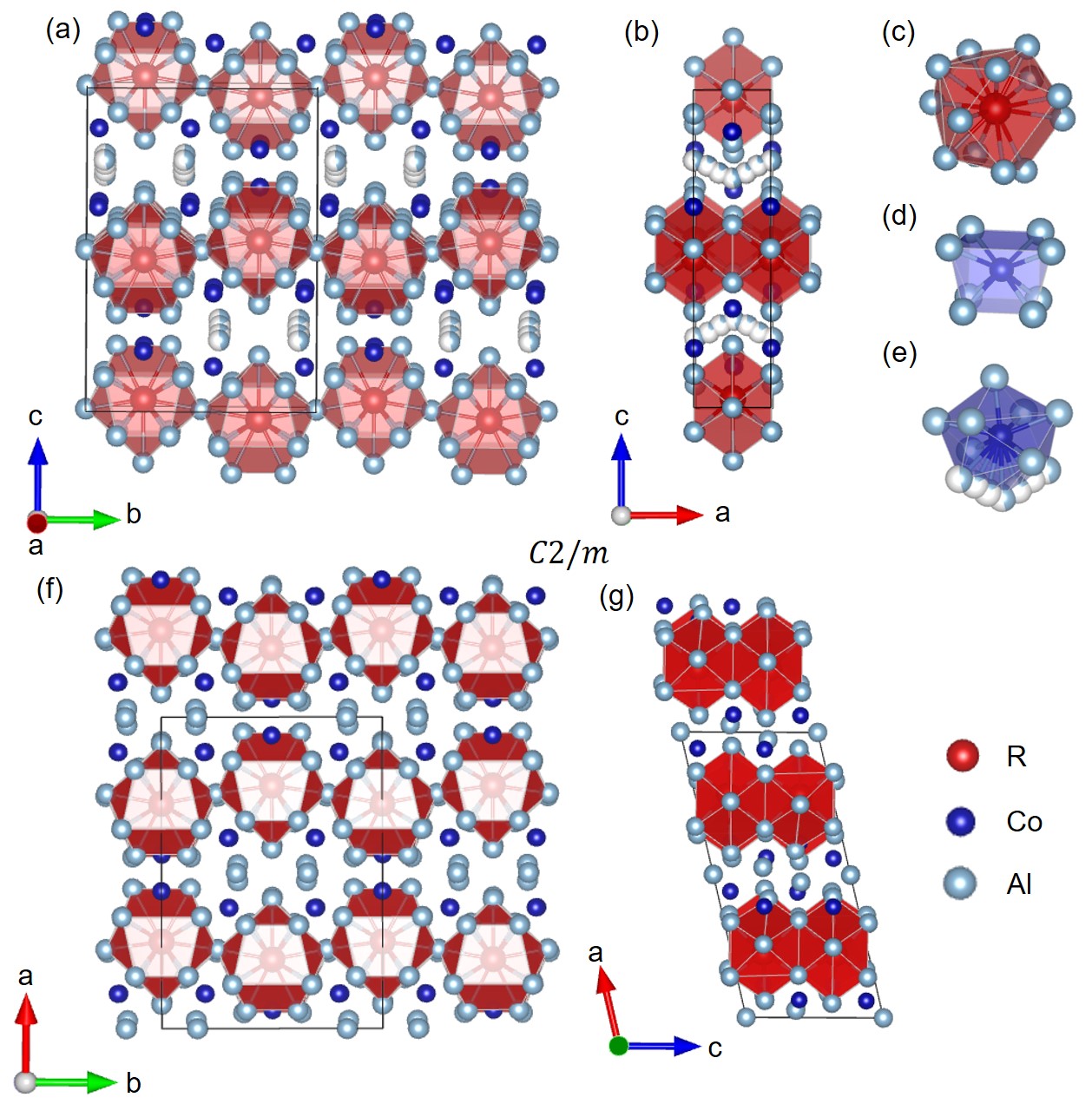} 
\caption{\footnotesize{Crystal structure of $R_{2}\text{Co}_{6}\text{Al}_{20-\delta}$. (a) shows the orthorhombic $Imma$ structure viewed down the a-axis and (b) the same structure viewed down the b-axis. Both views highlight the strongly disordered Al6 and Al7 atoms that fill the channels between R-Al polyhedra and run along the a-axis. (c-e) show the crystallograpically unique coordination polyhedra of the R (red) and Co (blue) atoms, with (e) emphasizing the disordered Al6 and Al7 atoms atoms.  (f) and (g) show the monoclinic $C 2/m$ crystal structure previously proposed for $R_{\text{2}}$Co$_6$Al$_{19}$ viewed down the $c-$ and $b-$axes, respectively. The two views emphasize the similarity between orthorhombic and monoclinic structures, with ordered the atomic coordinates Al1 with coordinates (0, 0.167, 0) and Al4 with coordinates (0.0258, 0.174, 0.365) atoms in the monoclinic structure and disordered Al chains in the orthorhombic arrangement. 
\protect\label{fig:crystal_structure}}}
\end{figure}

Single crystals of $R_{2}\text{Co}_{6}\text{Al}_{20-\delta}$ ($R$=Gd-Tm and Y) were grown using a self-flux method \cite{Canfield_2020} with an initial composition of 1:2:20 of $R$:Co:Al respectively. High purity elements (3N) were enclosed in Canfield Crucible Sets \citep{Canfield2016,LSPCeramics} under an $\approx$ 1/6 atmosphere
of Argon in a sealed amorphous silica ampule. The packed ampule was placed in a furnace, heated to 1180$\degree $C in 7 hours, held at 1180$\degree $C for 24 hours to ensure the material melted completely and then cooled gradually over 150 hours to 900$\degree $C, the temperature at which the excess flux was decanted from the single crystals using a centrifuge \citep{Canfield_2020}. The growths produced thin, rod-like crystals similar to the one shown in figure \ref{fig:crystal_image}.

In order to check phase purity/confirm the crystal structure of the grown crystals, powder X-Ray diffraction (PXRD) and single crystal X-ray diffraction (SCXRD) experiments were made. PXRD  measurements were made using a Rigaku MiniFlex II powder diffractometer with Cu-K$\alpha$ radiation ($\lambda$ = 1.5406 \AA). For each $R$-based material, a few crystals were finely ground and spread evenly on a single crystal Si, zero background holder, with the help of a small amount of vacuum grease. Intensities were collected for 2$\theta$ ranging from 5$\degree$ to 100$\degree$, in step sizes of 0.01$\degree$, counting for 4 seconds at each step. Also, Temperature dependent PXRD measurements were performed from 12 to 300 K using a Rigaku SmartLab X-ray diffractometer equipped with an Oxford K450 helium cryosystem and Cu K$\alpha$ radiation. Rietveld refinement was performed for each diffraction pattern using the GSASII software package \citep{von2014small}. 

Single crystal X-ray diffraction was performed using a Rigaku XtaLab Synergy-S diffractometer with Ag radiation ($\lambda$ = 0.56087 \AA), in transmission mode, operating at $65$ kV and $0.67$ mA. The samples were held on a nylon loop with Apiezon N grease. All measurements were performed at room temperature ($\approx$ 300 K). The total number of runs and images was based on the strategy calculation from the program CrysAlisPro (Rigaku OD, 2023). The data integration and reduction were also performed using CrysAlisPro, and a numerical absorption correction was applied based on Gaussian integration over a face-indexed crystal. The structures were solved by intrinsic phasing using the SHELXT software package and were refined with SHELXL.

 DC magnetization measurements were performed as a function of temperature (1.8 - 300 K) and field (up to 50 kOe) in Quantum Design SQUID magnetometers. The measurements were conducted on rod-shaped samples, where the long axis corresponds to crystallographic $a$-axis, as determined through Laue diffraction analysis. Field cooling (FC) and zero-field cooling (ZFC) procedures were used for all measurements, and the field was applied perpendicular to and along the $a$-axis, unless otherwise noted, temperature dependent magnetization was measured at 1 kOe for almost every sample, where for all $R{_2}$Co${_6}$Al${_{20-\delta}}$ samples, the magnetization is linear in field, allowing us to define the susceptibility as $\chi = M/H$. The exception was Ho$_2$Co${_6}$Al${_{20-\delta}}$ which had the lowest $T_N$, close to 1.8 K , for which we used 100 Oe for lower temperature M(T) data. To measure the magnetization along crystallographically significant directions, the samples had to be manually aligned. To hold the sample in place, it was placed between two half plastic straws which were inserted inside another plastic straw. We also measured the magnetic susceptibility of powdered samples contained inside a gelatin capsule. To eliminate the contribution from the sample holder, the gelatin capsule was first measured empty. To prepare the sample, we finely ground several single crystals to minimize preferential orientation effects. The resulting powder was placed in a gelatin capsule, lightly compacted using the capsule's upper part, and sealed with the cap inverted (a pinhole was also poked in the capsule to allow for evacuation). The prepared capsule was then secured inside a plastic straw, ensuring dimensions of the straw were appropriate to prevent any movement during the measurement process. The magnetic response of the empty capsule was subtracted from that of the capsule containing the sample to remove the background signal arising from the gelatin material.

Temperature-dependent specific heat measurements were carried out in a Quantum Design DynaCool PPMS using the relaxation technique as implemented in the heat capacity option, using heat pulses and fitting the temperature relaxation with a two- $\tau$ model. The addenda were first measured by applying a small amount of N-grease to the sample platform and recording its contribution. The sample was then mounted on the same grease layer and measured under identical conditions. The total heat capacity was automatically corrected for the addenda contribution using the instrument’s software, yielding the intrinsic heat capacity of the sample. Using the corrected heat-capacity data, the magnetic entropy for all compounds was calculated; a detailed description of the procedure used to obtain the magnetic entropy is provided in Appendix D.

Electrical resistivity of every member of the $R_{2}\text{Co}_{6}\text{Al}_{20-\delta}$ series was measured using the resistivity option of a Quantum Design Dynacool (PPMS) instrument with $8$ mA current and constant mode option for other parameters. The electrical resistivity was measured on a rod-like single crystal ($\approx 1.5$~mm $\times$ 0.5~mm $\times$ 0.2~mm) using a standard four-probe geometry, with the current applied along the crystallographic a-axis. The geometric uncertainty in the measurement is estimated to be within $\approx 10$\%. Electrical contacts with less than 1.5 $\Omega$ resistance were achieved by spot-welding 25 $\mu$m Pt wire to the samples, followed by adding Epotek H20E silver epoxy, and curing the latter for 1 hour at $120^{\circ}$C.

\label{experimental}

\section{Results and Discussion}
\subsection{Structure}

\begin{table}
\centering
\caption{Refined atomic positions, site occupancy, and isotropic thermal displacement parameters for Gd$_{2}$Co$_6$Al$_{20-\delta}$, $\delta\sim 1$, according to SCXRD.}

\small
\begin{tabular}{lcccccc}
\hline
\hline
At. & Site & $x$ & $y$ & $z$ & Occ. & $U_{\text
{iso}}$\\
\hline
Gd01 & $4e$ & 1 & 3/4 & 0.47001(2) & 1 & 0.00891(6) \\
Co01 & $4e$ & 0 & 1/4 & 0.81241(4) & 1 & 0.00497(11)\\
Co02 & $8h$ & 0 & 0.55226(4) & 0.63207(3) & 1 & 0.00649(9) \\
Al01 & $4b$ & 0 & 0.36691(11) & 0.69616(7) & 1 & 0.0122(3) \\
Al02 & $8h$ & 0 & 1/2 & 1/2 & 1 & 0.0122(3) \\
Al03 & $8h$ & 0 & 0.42080(12) & 0.89396(8) & 1 & 0.0113(2)\\
Al04 & $8h$ & 1/2 & 0.63579(11) & 0.57511(9) & 1 & 0.0127(2) \\
Al05 & $4e$ & 0 & 3/4 & 0.66511(11) & 1 & 0.0130(3) \\
Al06 & $8h$ & 0 & 0.5780(3) & 0.77787(19) & 0.446(6) & 0.0242(8) \\
Al07 & $16j$ & 0.167(3) & 0.5779(5) & 0.7625(3) & 0.164(4) & 0.0242(15) \\

\hline
\hline
\label{tab:scxrd_Gd} 
\end{tabular}
\end{table}
\medskip

\addtolength{\tabcolsep}{-3pt} 

The $R_{2}$Co$_{6}$Al$_{19}$ materials containing the heavy rare-earth elements $R$ = Gd-Tm,Y were first identified by Morozkin et al., who found traces of these compounds in arc-melted $R$-Co-Al mixtures \citep{R2Co6Al9poly}. On the basis of powder X-ray diffraction patterns, they suggested that these materials crystalize in the U$_{2}$Co$_{6}$Al$_{19}$ structure type with monoclinic $C2/m$ (\#12) symmetry, like the light $R$ members \citep{LaPr,Ce}. Here, we explored the crystal structures of our $R_{2}$Co$_{6}$Al$_{20-\delta}$  samples in more detail using single crystal X-ray diffraction.  

In contrast to the $C2/m$ symmetry suggested previously, our single crystal diffraction data indicates that the $R_{2}$Co$_{6}$Al$_{20-\delta}$ materials, with heavy $R$ = Gd-Tm, adopt a orthorhombic structure with space group $Imma$ (\#74). Our SCXRD data does not show peak splittings or superlattice peaks suggestive of a lower symmetry. The dataset could be satisfactorily refined in the higher-symmetry $Imma$ model, with approximately 99 $\%$ of the observed diffraction peaks indexed and no significant unaccounted reflections. To further address the crystal symmetry, figure \ref{fig:PXRD data of GdCoAl} shows a powder X-ray diffraction pattern collected on a Gd$_{2}$Co$_{6}$Al$_{20-\delta}$ sample. 
Whereas good quality fits to the powder data can be obtained with either $Imma$ or $C2/m$ structural models ($C2/m$ is a subgroup of $Imma$), the refinement in the $Imma$ structure yields slightly better statistics with R = 7.7 $\%$ compared to R = 8.9 $\%$ for refinement against the $C2/m$ structure. As shown in figure \ref{fig:PXRD data of GdCoAl}$c$, the theoretical pattern for the $C2/m$ is be expected to show several weak, low angle reflections that are forbidden in the body-centered $Imma$ group. Within the resolution of our measurement, figure \ref{fig:PXRD data of GdCoAl}$c$ demonstrates that we do not observe any of these $C2/m$ peaks. Taken together, our single and powder X-ray diffraction measurements support $Imma$ symmetry for the $R_{2}$Co$_{6}$Al$_{19}$  ($R$ = Gd-Tm) crystal structure.

Figures \ref{fig:crystal_structure}$a-e$ show several schematic views of the structure solution obtained from our SCXRD refinements, and the atomic coordinates, site occupancies, and thermal displacement parameters are listed in Table \ref{tab:scxrd_Gd}. Additional crystallographic information regarding the refinements are given in Table \ref{tab:scxrd_Gd}(Data is shown only for the Gd$_{2}$Co$_{6}$Al$_{20-\delta}$; results for the other compounds are provided in Appendix C.). The structure contains one crystallographically unique $R$ site (Wyckoff position 4e), in which the $R$ atoms sit in a mm2 (orthorhombic symmetry) site at the center of a 13-fold coordination polyhedron of Al atoms, highlighted in figure \ref{fig:crystal_structure}$c$. The extended structure of $R_{2}$Co$_{6}$Al$_{20-\delta}$ ($R$ = Gd-Tm) is formed by the three-dimensional packing of these $R$-centered Al polyhedra. As emphasized in figures \ref{fig:crystal_structure}$a$ and \ref{fig:crystal_structure}$b$ respectively, the $R$-Al polyhedral form corner sharing links along the b-axis and face sharing links along the a-axis. There are two unique Co sites. Figure \ref{fig:crystal_structure}$d$ shows that the Co1 atoms are coordinated by eight Al atoms with a distorted rectangular prismatic geometry. The second, Co2, sites are nine-fold coordinated by Al atoms and sit at the center of a distorted tri-capped trigonal prism. Both Co-Al coordination polyhedra form face-sharing linkages along the a-axis. Notably, the Co2 atoms are coordinated to two heavily disordered, Al6 and Al7 sites.  Attempts to assign these two Al atoms a fixed position resulted in unphysically large thermal displacement parameters, and we found a reasonable refinement could be achieved by assuming these atoms to have site occupancy factors substantially lower than 1, as listed in Table \ref{tab:CrystalSturcture_Information} Appendix C. Given the otherwise unrealistically close contacts between Al6 and Al7 atoms ($\approx$1.5 \AA), the physical interpretation of this result is that the Al6 and Al7 atoms do not have a well-defined position, but randomly occupy the voids between $R$-Al and Co-Al polyhedra in a zig-zag chain arrangement most clearly illustrated in figure \ref{fig:crystal_structure}$b$.

It is especially interesting to compare the orthorhombic, $Imma$, structure described above with the monoclinic, $C2/m$ structure, adopted by the light $R$ = La-Sm members of the $R_{2}$Co$_{6}$Al$_{19}$ series. Figures \ref{fig:crystal_structure}$f$ and \ref{fig:crystal_structure}$g$ show the $C2/m$ U$_{2}$Co$_{6}$Al$_{19}$-type structure viewed down the $c$- and $b$-axis respectively. It is clear that the salient motifs and local coordination geometries are identical in both crystal structures. The primary difference is that in the $C2/m$ U$_{2}$Co$_{6}$Al$_{19}$-type structure, the Al atoms are no longer disordered. Instead of the zig-zag like chains of random occupancy found in the $Imma$ arrangement, the respective Al atoms now have well defined crystallographic positions, which results in a lower symmetry associated with a change in centering and doubling of the unit cell.

\subsection{Physical Properties}

We now present physical properties for each of the $R$-based materials. In figure 18 of Appendix A, we show a picture of one of our single crystals, which display a representative morphology. However, for most members of the family, the crystals were significantly smaller. In particular, their reduced area perpendicular to the rod axis posed a challenge for orienting the crystals along specific in-plane directions. For this reason, the magnetic anisotropy was characterized by magnetization measurements with the field parallel to the $a$-axis and perpendicular to the $a$-axis. Moreover, we compare the magnetic measurements of the powdered and polycrystalline average obtained from Ho$_{2}$Co$_{6}$Al$_{20-\delta}$ single-crystal, and this result motivated our magnetization measurements of powdered samples (detailed explanation in Appendix A).

\subsubsection{Y$_{2}$Co$_{6}$Al$_{20-\delta}$}

\begin{figure}
\centering{}\includegraphics[width=0.45\textwidth]{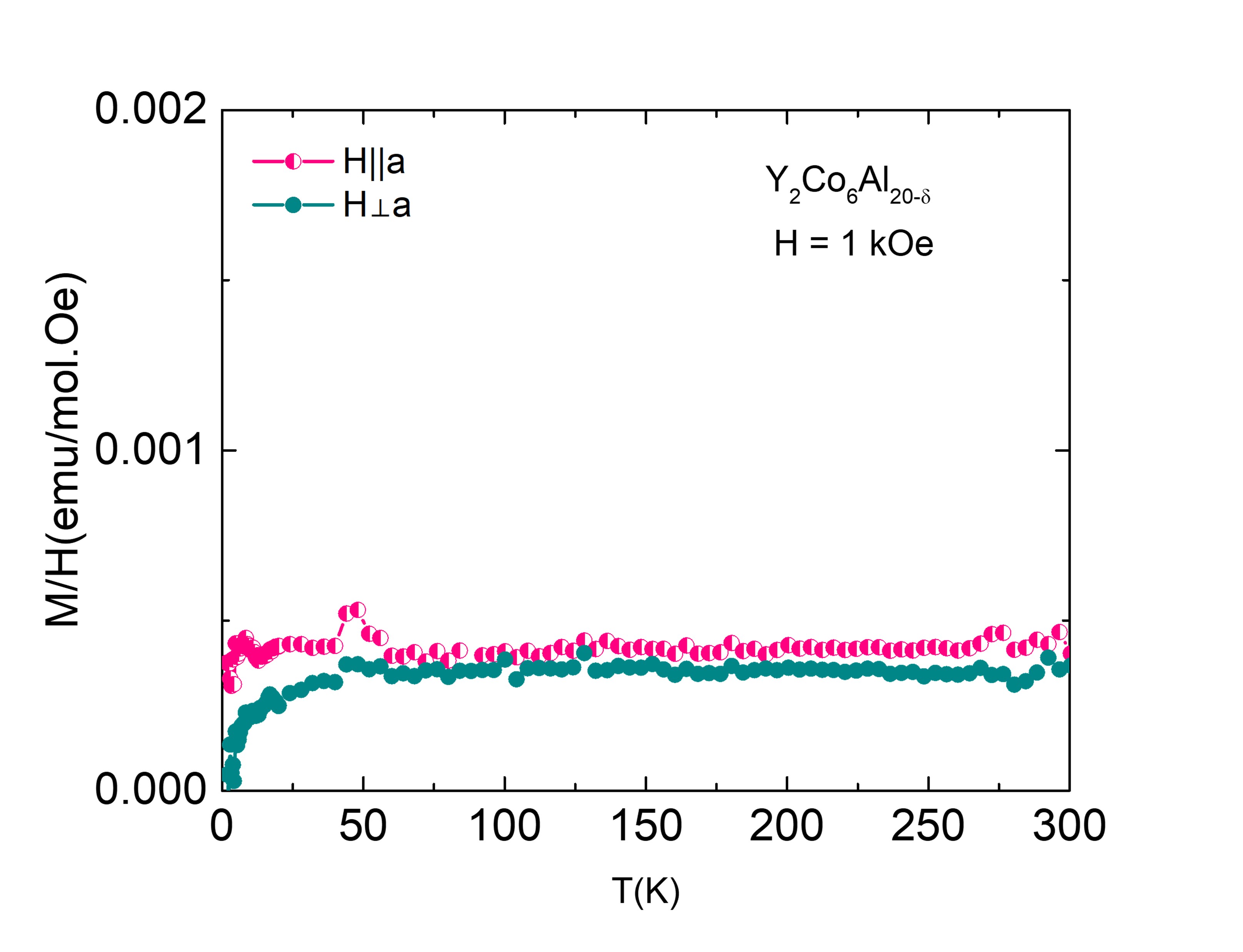} 
\caption{\footnotesize{ Temperature dependent magnetization of Y$_{2}$Co$_{6}$Al$_{20-\delta}$ for both direction ($H\parallel a$ and $H\perp a$). 
\protect\label{fig:YCoAl magnetization}}}
\end{figure}

\begin{figure}
\centering{}\includegraphics[width=0.45\textwidth]{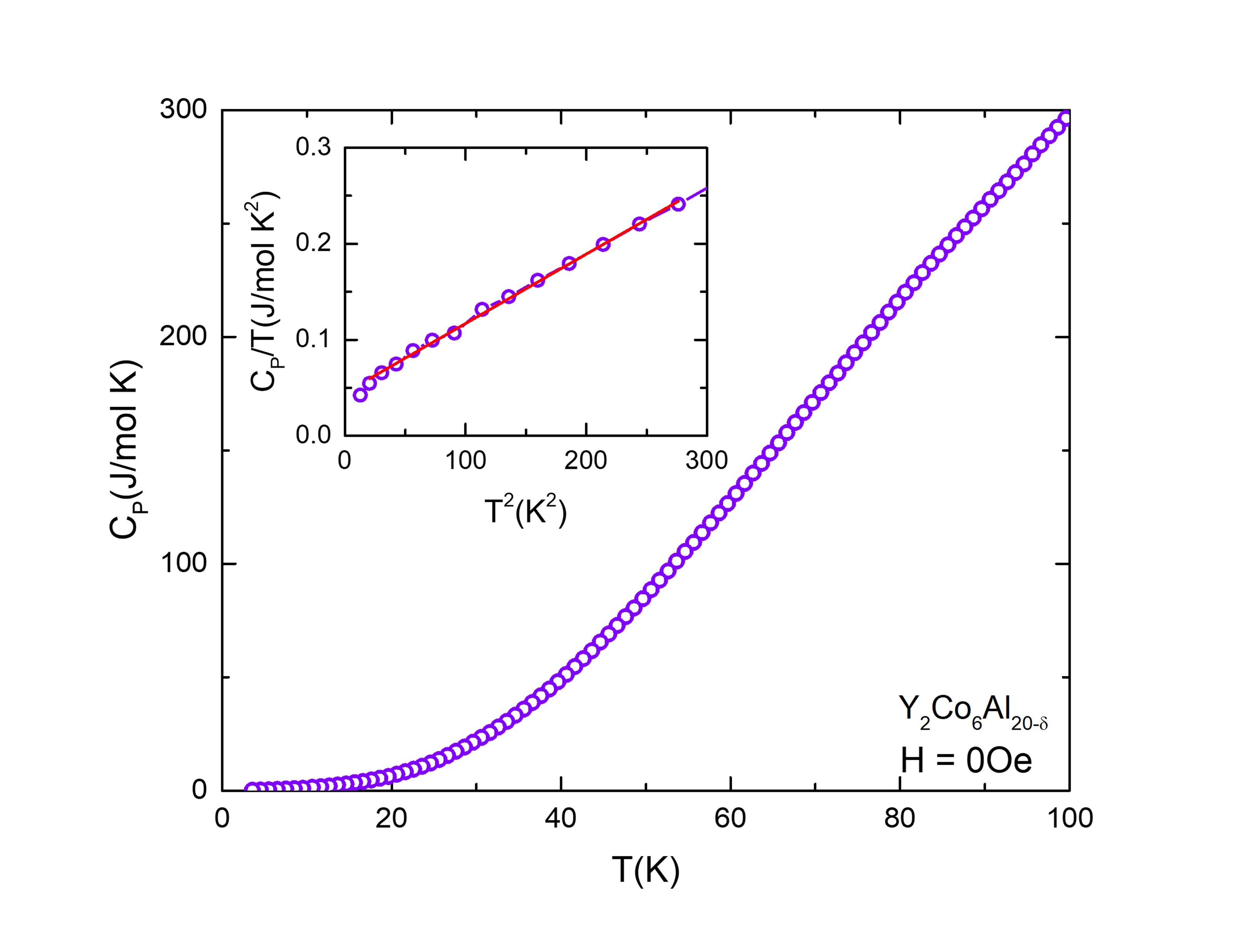} 
\caption{\footnotesize{ Temperature dependent specific heat measurements of Y$_{2}$Co$_{6}$Al$_{20-\delta}$ from 1.8 to 100 K range, in inset we are showing $C_P$/T versus $T^2$. 
\protect\label{fig:YCoAl_Cp}}}
\end{figure}

\begin{figure} 
\centering{}\includegraphics[width=0.45\textwidth]{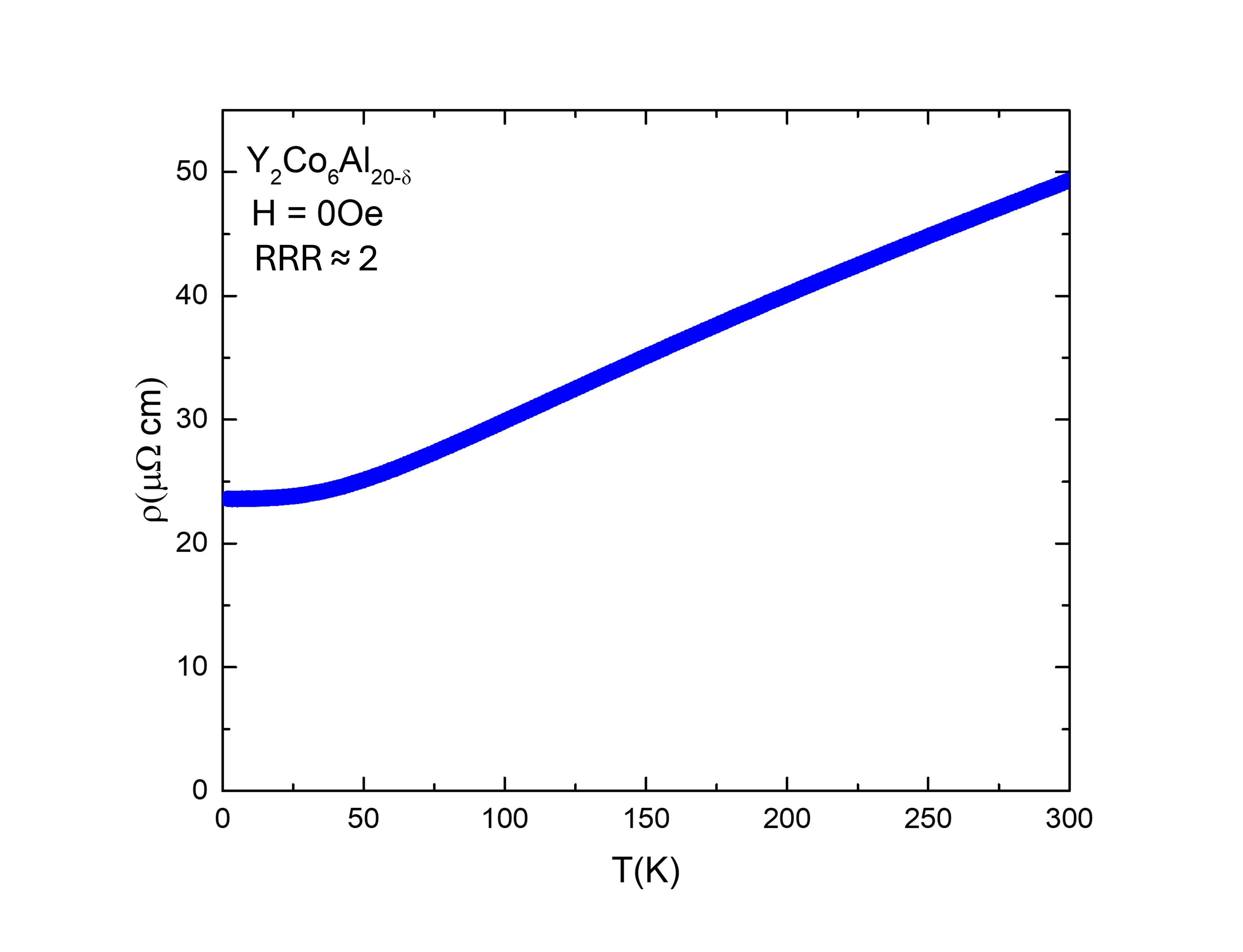} 
\caption{\footnotesize{Temperature dependent resistivity measurements of Y$_{2}$Co$_{6}$Al$_{20-\delta}$ . 
\protect\label{fig:YCoAl_R}}}
\end{figure}

 Temperature-dependent magnetization is shown in figure \ref{fig:YCoAl magnetization} for single crystalline Y$_{2}$Co$_{6}$Al$_{20-\delta}$ for both crystallographic directions; it exhibits temperature-independent Pauli-like paramagnetism, consistent with the non-moment-bearing nature of the Y$^{3+}$ ion. The data shows no indication of any phase transition across the measured temperature range however, we see down turn at low temperature that is coming from inaccuracies associated with the background subtraction. It should be noted that these data strongly indicate that the Co is non-moment-bearing in this series.The specific heat data shown in figure \ref{fig:YCoAl_Cp} also shows no indication of any phase transitions and will be used as a reference to approximate the sum of electronic and phonon contributions to the specific heat in $R$$_{2}$Co$_{6}$Al$_{20-\delta}$ to calculate magnetic entropy. In the inset of figure \ref{fig:YCoAl_Cp}, the low-temperature specific heat data are plotted as  $C_P/T $ versus $T^2$. The low temperature data were fitted using the expression $ \frac{C_P}{T} = \gamma + \beta T^2$,
 where the linear term $\gamma$ represents the electronic contribution and the $\beta T^2$ term corresponds to the phonon contribution. From the fit, we obtained a Sommerfeld coefficient $\gamma$ = 45(2) {mJ mol$_{f.u.}^{-1}$ K$^{-2}$, and a Debye temperature $\Theta_D$ = 419(2)
 {K}, derived from the $\beta$ parameter. 

The electrical resistivity as a function of temperature, shown in figure \ref{fig:YCoAl_R} over the range 1.8–300 K, reveals metallic behavior with a residual resistivity ratio (RRR = R(300 K)/R(1.8 K)) of approximately 2. The relatively low RRR value likely reflects enhanced charge-carrier scattering arising from Al-site disorder associated with partial occupancy of the Al sublattice.  The resistivity data for Y$_{2}$Co$_{6}$Al$_{20-\delta}$ manifest no sharp features and is also consistent with the absence of any phase transitions between 1.8 and 300 K.

\subsubsection{Gd$_{2}$Co$_{6}$Al$_{20-\delta}$}

\begin{figure*}
\centering
\includegraphics[width=\linewidth]{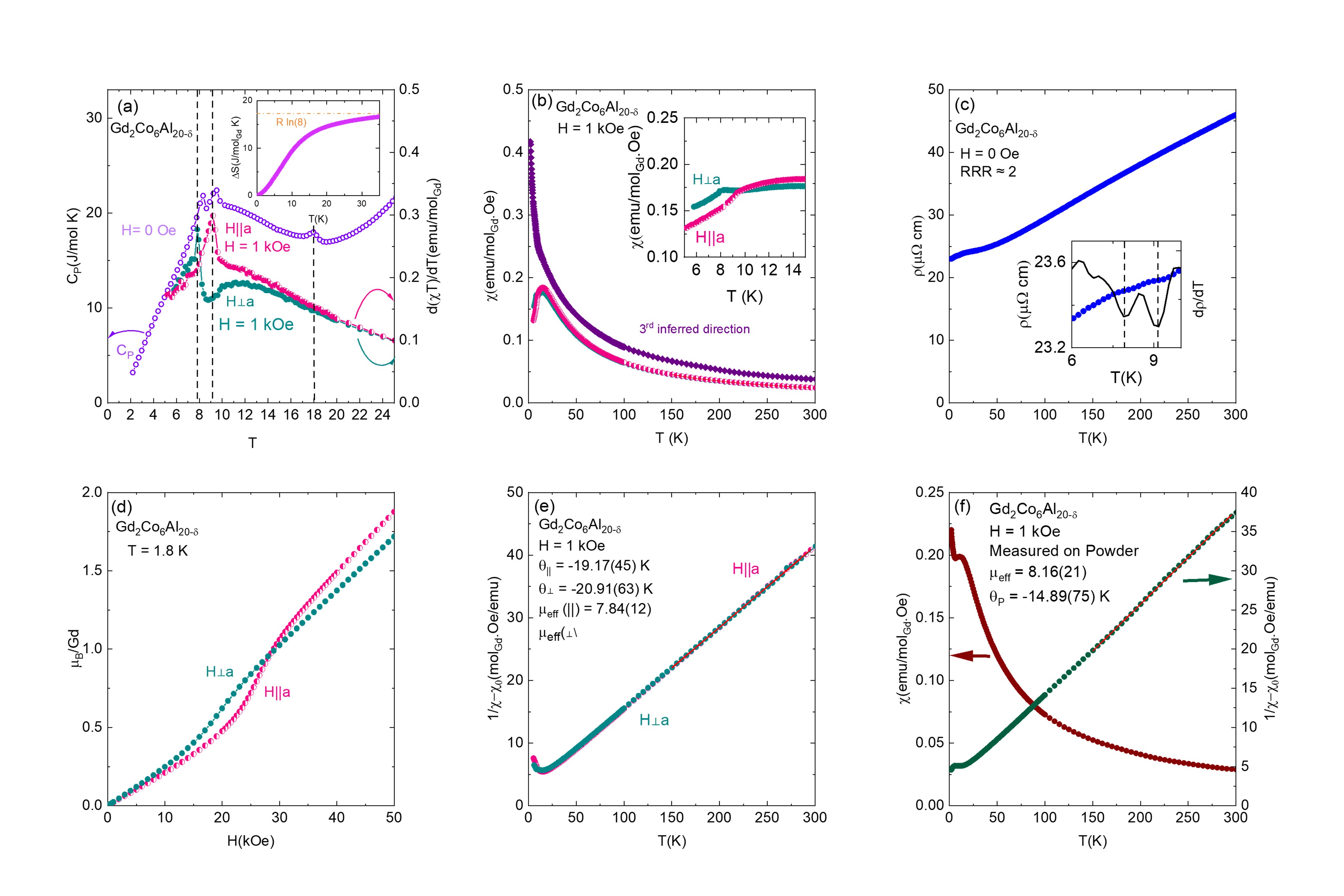} 
\caption {\footnotesize{Physical properties of Gd$_{2}$Co$_{6}$Al$_{20-\delta}$ single crystal. $(a)$ $C_{p}$ (left axis) and $d(\chi T)/dT$ (right-axis) obtained
for $H\parallel a$ ( half open symbols ) and $H\perp a$ ( solid
symbols). The data clearly suggest two AFM transitions $\approx$ $9.7$ K and $\approx 8.2$K, inset: estimation of magnetic entropy (see text for details). $(b)$ Temperature dependent susceptibility data was measured for both direction($\chi_\parallel$ and $\chi_\perp$) and inset shows in detail the $T$-region close to the AFM transitions. $(c)$ Temperature dependent electrical resistivity measurement (I$||a$). In the inset, we present $\rho$ and $d\rho/dT$ in the region close to the AFM transitions(dashed line). The transitions, $T_{N}$ and $T_{SR}$, can be observed as the dips in temperature derivative. $(d)$ Isothermal magnetization measurements at $T=1.8$ K for applied field up to $5$ T for $H\parallel a$ ( half open symbols ) and $H\perp a$ ( solid symbols). $(e)$ $\chi^{-1}$ data obtained for both $H\parallel a$ and $H\perp a$ and used to extract parameters $\theta_{\text{||}}$ and $\theta_{\perp}$. $(f)$ $\chi_{\text{powder}}$ (left-axis) and $(\chi_{\text{powder}}-\chi_{0})^{-1}$ (right-axis) data. The solid lines in figure $(e)$ and $(f)$ show the Curie Weiss fit used to estimate $\mu_{\text{eff}}$ and $\theta$.}}
\label{fig:Gd_based_A=000020}
\end{figure*}

\begin{figure}
\centering
\includegraphics[width=\linewidth]{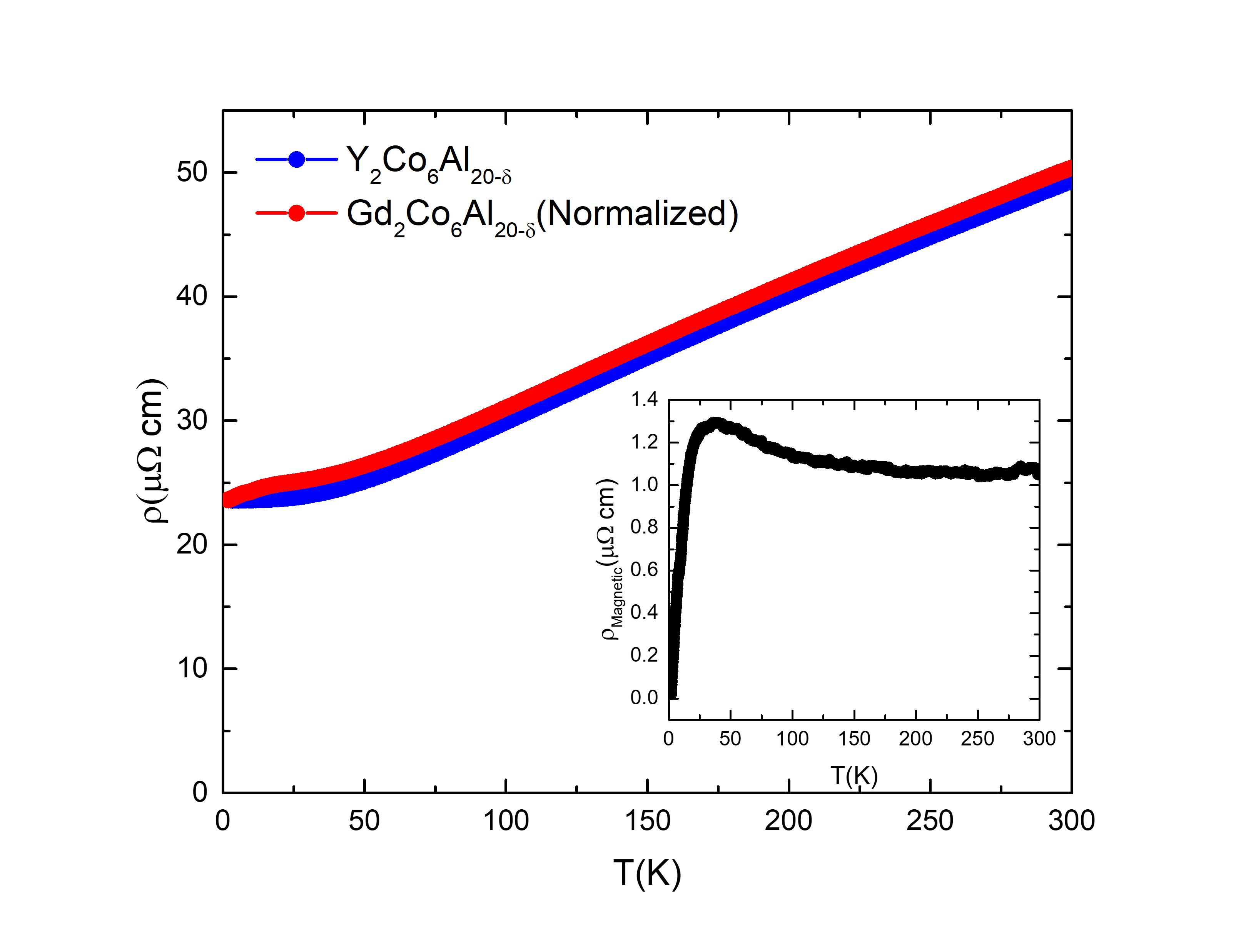} 
\caption {\footnotesize{Temperature dependence of the magnetic contribution to the electrical resistivity, obtained by subtracting the resistivity of the nonmagnetic Y$_{2}$Co$_{6}$Al$_{20-\delta}$ from that of the Gd$_{2}$Co$_{6}$Al$_{20-\delta}$.}}
\label{fig:R_mag}
\end{figure}

\begin{figure}
\centering
\includegraphics[width= 1.1\linewidth]{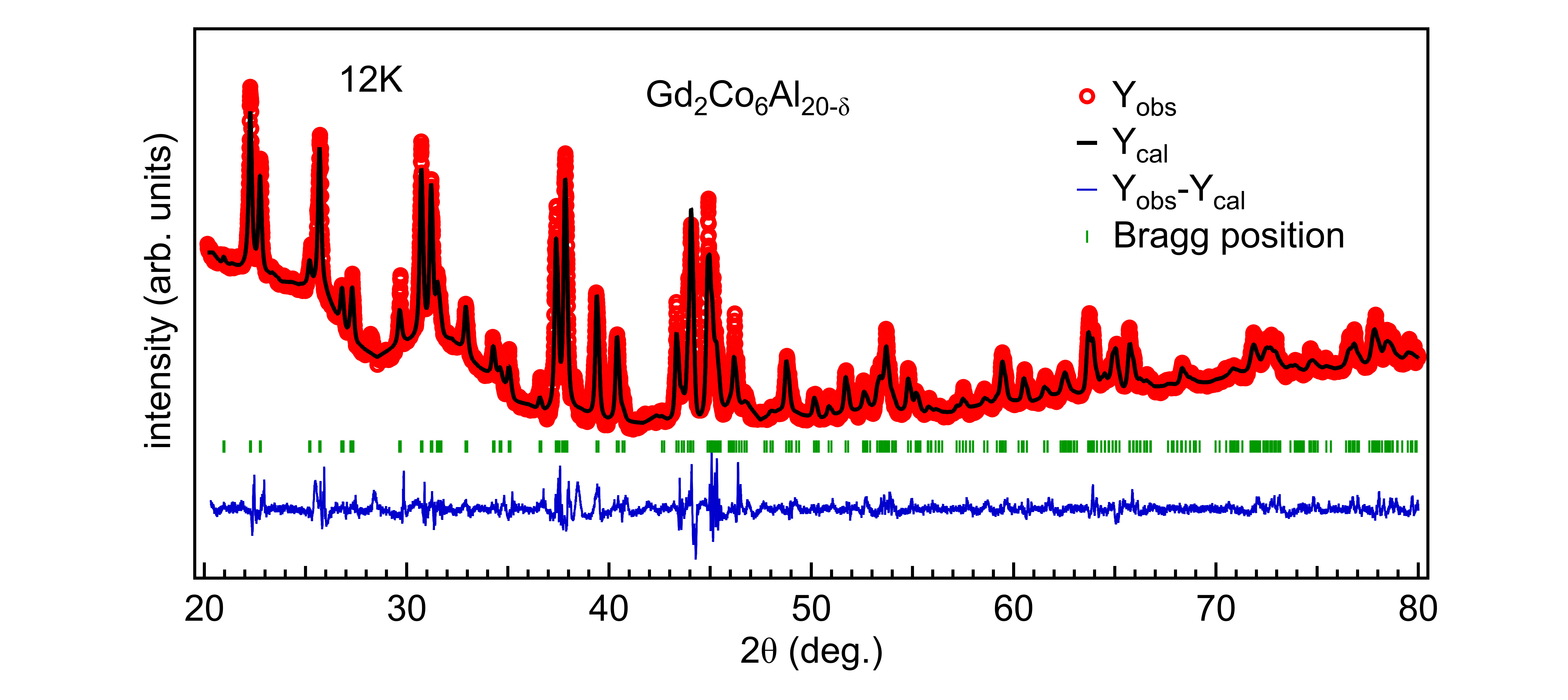} 
\caption {\footnotesize{ Powder X-ray diffraction (PXRD) data collected of Gd$_{2}$Co$_{6}$Al$_{20-\delta}$ at 12 K, including the Rietveld refinement using the $Imma$ space group.}}
\label{fig:12K}
\end{figure}

\begin{figure}
\centering
\includegraphics[width=\linewidth]{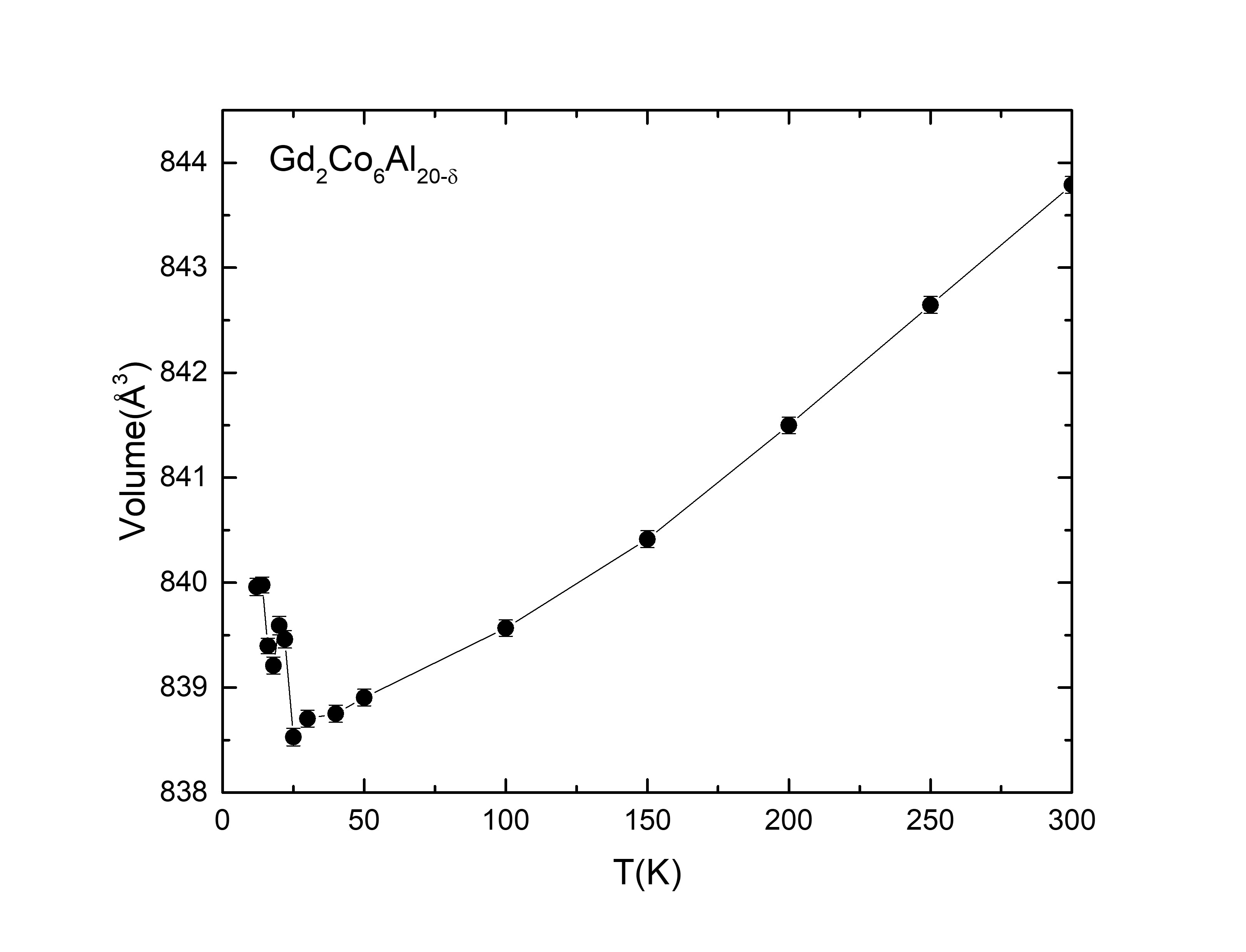} 
\caption {\footnotesize{Temperature dependence of the unit-cell volume obtained from temperature-dependent PXRD. The data show an anomalous deviation from conventional thermal contraction around 25 K, where the unit-cell volume exhibits a increase upon cooling.}}
\label{fig:Volume_T}
\end{figure}

Magnetic properties of the Gd$_{2}$Co$_{6}$Al$_{20-\delta}$ single crystal stems from the $4f^{7}$ configuration  of the Gd$^{3+}$ ion, for which $J = 7/2$ due to $L = 0$ and $S = 7/2$. Since $L=0$, CEF splitting of the Hund's rule ground state effects is absent. In figures \ref{fig:Gd_based_A=000020}$a$-$c$, we present the specific heat ($C_{p}$), magnetic susceptibility ($\chi$) and resistivity ($\rho$) data as a function of temperature ($T$) of a Gd$_{2}$Co$_{6}$Al$_{20-\delta}$ single crystal.


The $C_p$(T) data (left-axis of fig. \ref{fig:Gd_based_A=000020}$a$) was measured in the absence of an applied magnetic field ($H$). The $C_p$(T) exhibits three distinct sharp peaks at temperatures $ 9.7(3)$ K, $8.2(3) $ K and around $19$ K. However, the peaks at temperatures $ 9.7(3)$ K and $8.2(3) $ K are sharp, whereas the one at around $19$ K is a broad bump. The criteria used for determining the transition temperatures is discussed in appendix B and applied to each transition temperature for each member of the series. We begin by discussing the two lower-temperature transitions, while the higher-temperature transition will be discussed subsequently.

As expected, the temperature-dependent magnetic susceptibility for the single crystal of the Gd-based compound remains nearly isotropic in the paramagnetic region (Fig. \ref{fig:Gd_based_A=000020}$b$). At lower temperature both $\chi_{\parallel}$ (field applied along $a$ - axis) and $\chi_{\perp}$ (field applied perpendicular to $a$ - axis) show features consistent with antiferromagnetic (AFM) ordering, evidenced by the peak-like shape of $\chi(T)$. Magnetic susceptibility was measured along two orthogonal crystallographic directions, as well as in the powder-averaged measurements, which is taken to approximate the average for all directions. The susceptibility along the third direction was inferred using equation \ref{eqn:X}.
\begin{equation} \label{eqn:X}
  \chi_{3^{rd}\text{inferred}} = 3*\chi_{\text{powder,avg}} - \chi_{\perp} - \chi_{\parallel}.  
\end{equation}

 This method was applied consistently across all compounds. Nevertheless, in several cases, evidence of preferential orientation is apparent.
 
The data in figure \ref{fig:Gd_based_A=000020}$b$ was used to compute $\frac{d(\chi T)}{dT}$, which is expected to scale with $C_p$ close to a second-order AFM transition \citep{fisher_relation_1962}. In figure \ref{fig:Gd_based_A=000020}$a$ (right axis) two distinct peaks are observed for $\chi_{\parallel}$ and $\chi_{\perp}$ at $\approx 9.5$ K and $\approx 8.2$ K, respectively. These peaks coincide (within the uncertainty) with those observed in $C_p$, where the peak observed for $\chi_{\parallel}$   matches the $9.7$ K temperature peak in $C_p$, and the peak observed for $\chi_{\perp}$ matches the $\approx 8.2$ K temperature peak in $C_p$. We identify $T_N \approx 9.7$ K as the antiferromagnetic ordering temperature, whereas $T_{SR} \approx 8.2$ K  is likely to be a spin reorientation transition. 

In figure \ref{fig:Gd_based_A=000020}$e$, the inverse susceptibilities, $\chi_{\parallel}^{-1}$ and $\chi_{\perp}^{-1}$, derived from the data presented in figure \ref{fig:Gd_based_A=000020}$b$, are shown. Furthermore, the susceptibility measured on powder samples, $\chi_{\text{powder}}$ and its inverse($\chi_{\text{powder}}^{-1}$) are displayed in figure \ref{fig:Gd_based_A=000020}$f$.  As shown in figure \ref{fig:Gd_based_A=000020}$e$ and \ref{fig:Gd_based_A=000020}$f$, the data follows a typical Curie-Weiss behavior and is nearly isotropic down to $\approx 25\,\text{K}$, below which magnetic anisotropy becomes apparent. In the absence of crystal electric field (CEF) effects, this observed anisotropy must be attributed to the anisotropy of the exchange interactions, which also likely drives the emergence of the two distinct magnetic transitions. This anisotropic exchange interaction appears to be a common characteristic of all $R$-based compounds in this series. However, its effects are most pronounced in Gd$_2$Co$_6$Al$_{20-\delta}$, where the absence of strong CEF-induced anisotropy allows the exchange anisotropy to dominate. $\chi_{\parallel}$, $\chi_{\perp}$ and $\chi_{powder}$  behave according to equation \ref{eqn:curie_R}  at higher temperature.

  \begin{equation}\label{eqn:curie_R}
    \chi=\frac{C}{(T-\theta)} - \chi_0
  \end{equation},

where $\chi_0$ represents the temperature-independent contribution to the magnetic susceptibility, accounting for non-Curie-Weiss components such as Pauli paramagnetism, Landau diamagnetism, and core diamagnetism; and $\theta$ is Curie-Weiss temperature.

In figure \ref{fig:Gd_based_A=000020}$e$, we estimated the magnetic anisotropy by fitting (shown with solid line in the figure) $\chi_{||}^{-1}$ and $\chi_{\perp}^{-1}$ data above $T=150$ K which, according to equation \ref{eqn:curie_R}, should be linear. We obtain $\theta_{||} \approx - 19$ 
K and $\theta_{\perp} \approx - 21$ K, where the negative values indicate Antiferromagnetic (AFM) correlations between local moments. The comparable values for $\theta_{||}$ and $\theta_{\perp}$ are consistent with the expectation for isotropic Gd moments. The effective  moment, $\mu_{\text{eff}}$, is determined by the  Curie-Weiss fit of $(\chi_{powder}-\chi_{0})^{-1}$ ($T >150$ K). The obtained value, $\mu_{\text{eff}} = 8.16(21)$ $\mu_{\text{B}}$/Gd, is in good agreement with the theoretical expected for Gd$^{3+}$ ($7.94$ $\mu_{\text{B}}$). 

We now turn to the electrical resistivity, $\rho$, as a function of temperature $T$ (Fig. \ref{fig:Gd_based_A=000020}$c$) . In the inset, we show $\rho$ (left-axis) and $d\rho/dT$ (right-axis) in a limited range of ordering temperatures. The lower temperature transition determined from \( C_p \) correspond to the two local minima observed in \( \frac{d\rho}{dT} \), as indicated by the dashed lines.

The isothermal magnetization at T = 1.8 K was obtained for both directions of the applied magnetic field: parallel ($H\parallel a$) and perpendicular ($H\perp a$) to the a-axis(Fig. \ref{fig:Gd_based_A=000020}$d$). In the field range of our measurements ($0 \leq H \leq 50$ kOe), the magnetization is far from the saturation value of $7 \mu_B $ expected for a $S=7/2$ spin system. A broad step-like feature is seen around $30$ kOe for $H\parallel a$, characteristic of a meta-magnetic transition. Also, we see a subtle increase for $H\perp a$ around $20$ kOe. 

Having discussed the lower temperature transitions, we now discuss the third transition observed around 18 K in the heat capacity ($C_p$), no distinct, corresponding anomalies are detected in either the magnetization or electrical resistivity. In order to better understand what may be associated with the $C_p$ feature near $18$ K, we measured temperature dependent X-ray diffraction data. We observe that overall crystal symmetry of the sample remains unchanged ($Imma$) down to 12 K, as shown in figure \ref{fig:12K}. However, interestingly, all the lattice parameters (see Fig. \ref{fig: Lattice} in the Appendix and the discussion therein) and the unit-cell volume, presented in figure \ref{fig:Volume_T}, show an upturn below $\approx$25 K that is inconsistent with any form or trivial thermal contraction and instead strongly suggests some form of structural phase transition. Although this temperature is slightly higher than the observed anomaly in the $C_p$(T) curves, a subtle change in the local crystal structure could be responsible for this behavior. Whereas we have not been able to infer a change in unit cell symmetry or basis with our current data, further investigations using single crystal x-ray or neutron diffraction are necessary to disentangle purely structural changes from possible magnetostriction effects in the present case.

 In the inset of figure \ref{fig:Gd_based_A=000020}$a$, we have shown the estimated magnetic entropy for Gd$_{2}$Co$_{6}$Al$_{20-\delta}$ (detailed explanation is discussed in Appendix E about entropy calculation). We would expect the majority of the magnetic entropy associated with the Gd$^{3+}$ (S = 7/2) moments to be recovered near the highest confirmed magnetic transition at $T_N \approx 9.7$ K. Instead, only approximately half of the expected (R$\ln(2S+1))$ entropy is recovered up to this temperature. The origin of the reduced magnetic entropy remains an open question and likely arises from several factor. The reduced magnetic entropy may be linked to the presence of magnetic fluctuations above ordering temperature. The magnetic contribution to the resistivity, obtained by subtracting the nonmagnetic Y-based compound, exhibits an upturn around 75 K and follow broad maxima at low temperature (Fig. \ref{fig:R_mag}), indicative of enhanced magnetic scattering. This behavior suggests the presence of magnetic fluctuations extending above the magnetic ordering temperature, which can possibly contributing in reduced entropy. Also, the precise nature of this structural change remains unresolved, any lattice change or symmetry change could contribute to the observed reduction in magnetic entropy. Additionally, the quasi-one-dimensional structural motif of this system may play a significant role, as reduced dimensionality. Similar behavior has been reported in insulating Gd-based quasi-one-dimensional spin-chain compounds, where strong intrachain correlations lead to broad entropy release and reduced entropy at the long-range ordering temperature, reflecting pronounced low-dimensional spin fluctuations \citep{doi:10.1021/acs.chemmater.3c03217}. In the $R$$_{2}$Co$_{6}$Al$_{20-\delta}$ compounds, the crystal structure contains chains of rare-earth ions extending along the $a$-axis, with an intrachain Gd–Gd distance of approximately 4.1215(3) Å and a substantially larger inter-chain separation of about 8.2430(5) Å. This separation of length scales is consistent with dominant intrachain interactions and reduced effective dimensionality. Nevertheless, additional measurements are required to definitively establish whether the transition near $18$ K is non-magnetic and to further substantiate the interpretation of quasi-one-dimensional magnetic correlations in this metallic system.

\subsubsection{$\text{Tb}_{2}\text{Co}_{6}\text{Al}_{20-\delta}$}

\begin{figure*}
\centering{}\includegraphics[width=\linewidth]{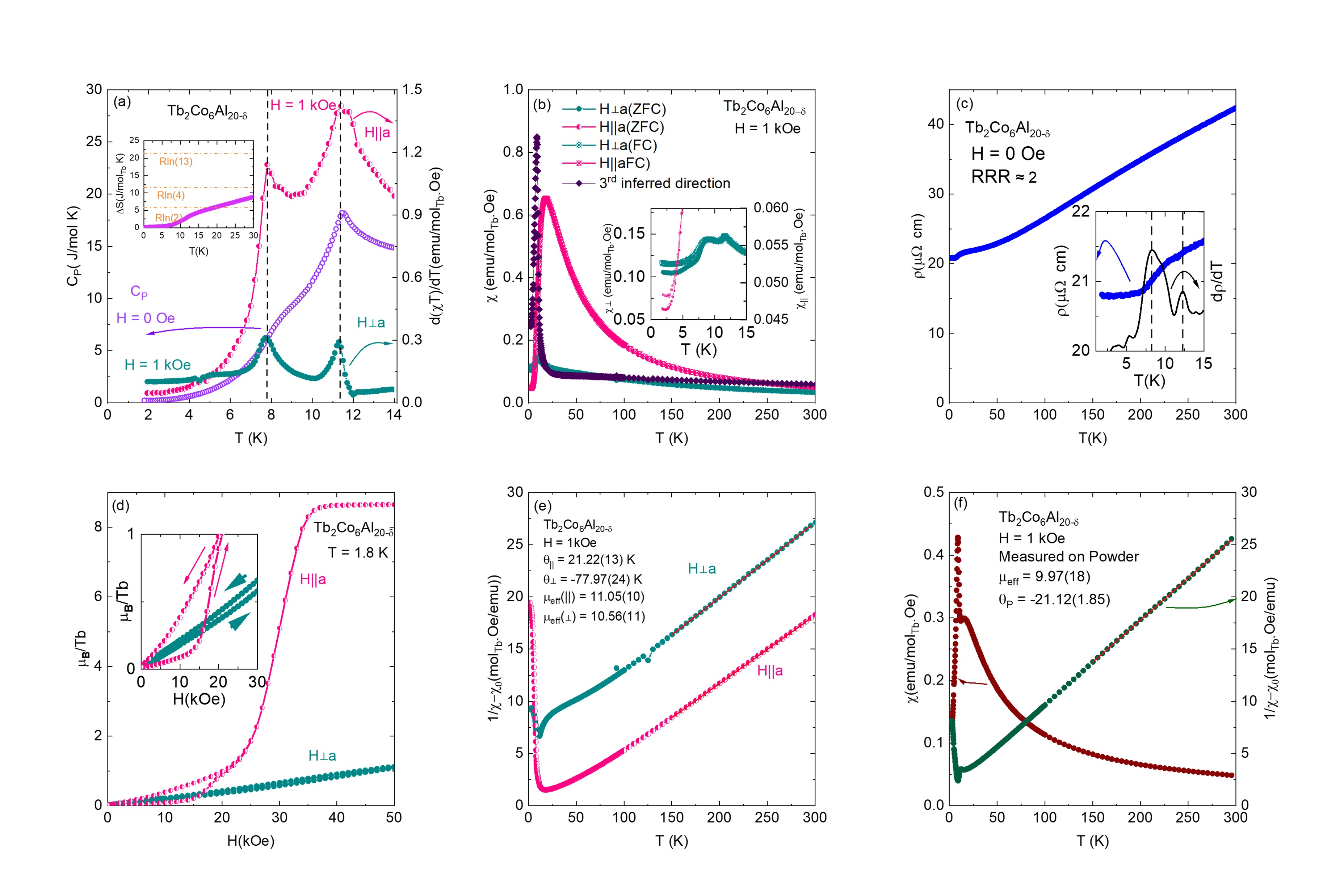} \caption{\footnotesize Physical properties of single crystalline {Tb$_{2}$Co$_{6}$Al$_{20-\delta}$ single crystal.
$(a)$ $C_{p}$(T) (left axis) and $d(\chi T)/dT$ (right-axis) obtained for $H\parallel a$ ( half open symbols ) and $H\perp a$ ( solid symbols). The $d(T\chi)/dT$ data (for both direction ) clearly suggest two AFM transitions that coincide well with features seen in the Cp(T) data, inset: magnetic entropy  inferred from the $C_{p}$(T) data (see text for details).. $(b)$ $\chi_{||}$ and $\chi_{\perp}$ was measured at $1$ kOe field. The inset of this figure shows in detail the $T$-region close to the AFM transitions for $\chi_{\perp}$ and ZFC and FC splitting for both directions. $(c)$ Temperature dependent electrical resistivity measurements (I $||a$).  In the inset, we present $\rho$ and $d\rho/dT$ in the range of the AFM transitions. The $T_{N}$ and $T_{{SR}}$ transition temperatures are marked by dashed lines. $(d)$ Isothermal magnetization measurements at $T=1.8$ K for applied field up to $5$ T for $H\parallel a$ ( half open symbols ) and $H\perp a$ ( solid symbols). $(e)$ Inverse susceptibility data obtained for $H\parallel a$ and $H\perp a$. $\theta_{\text{||}}$ and $\theta_{\perp}$ parameters are obtained from inverse susceptibility. $(f)$ $\chi_{\text{powder}}$ (left-axis) and $(\chi_{\text{powder}}-\chi_{0})^{-1}$ (right-axis) data used to extract $\mu_{\text{eff}}$, and the solid lines in figure $(e)$ and $(f)$ show the Curie-Weiss fit used to estimate $\mu_{\text{eff}}$ and $\theta$.
\protect\label{fig:Tb_based_A=000020}}}
\end{figure*}

For Tb$_{2}$Co$_{6}$Al$_{20-\delta}$, the specific heat measurements (Fig. \ref{fig:Tb_based_A=000020}$a$) exhibit a sharp peak at $11.8(3) $ K, indicating a phase transition between an ordered and a disordered state. At lower temperatures a broad curvature is observed around $8$ K. In the inset of figure \ref{fig:Tb_based_A=000020}$a$, we are showing magnetic entropy estimation, which close to R$ln 2$ around 20 K given the qualitative nature of the low temperature extrapolation of Cp data. However, the entropy continues to increase above this temperature, consistent with the higher lying CEF levels.

We show the temperature dependent $\chi_{\parallel}$ and $\chi_{\perp}$ in figure \ref{fig:Tb_based_A=000020}$b$. These curves show characteristics of typical AFM ordering, as we can see in the figure(peak like behavior in $\chi(T)$). $\frac{d(\chi T)}{dT}$ (right axis of fig. \ref{fig:Tb_based_A=000020}$a$ ) shows distinct peaks from both  $\chi_{\parallel}$ and $\chi_{\perp}$, at temperatures of $T_N \approx 11.7$ K, $T_{SR} \approx 8.0$ K. These transition temperatures are in good agreement, within the uncertainty, with those observed in $C_p$. The first transition, occurring at $T_N$, corresponds to the onset of long-range antiferromagnetic order, while the second, at $T_{SR}$, is likely associated with a spin reorientation may be similar to that observed in Gd based compound. In the inset of figure \ref{fig:Tb_based_A=000020}$b$, the $\chi_{\perp}$ data shows a small but resolvable bifurcation between ZFC and FC measurements at low temperatures. This splitting may be related to domain effects. At higher temperatures, the data transition to more uniform behavior, exhibiting typical Curie-Weiss behavior for both $\chi_{\parallel}$ and $\chi_{\perp}$, while still reflecting the CEF-induced anisotropy. Also, The susceptibility along the third direction was inferred using the same methodology applied to the Gd-based compound. As shown in figure \ref{fig:Tb_based_A=000020}$b$, the inferred values are higher than both measured orientations ($\chi_{\parallel}$ and $\chi_{\perp}$). Given that magnetization data in figure \ref{fig:Tb_based_A=000020}$d$ indicates that the $a$-axis is highly likely the magnetic easy axis, this anomalous result likely arises from strong preferential orientation within the powder sample. Instead of a truly random distribution, the crystallites likely aligned such that the $a$-axis contribution is over-represented in the experimental powder data, leading to an artificially high inferred value.

Temperature dependent resistivity ( Fig. \ref{fig:Tb_based_A=000020}$c$) shows a change in slope at low temperature. In the inset, we show $\rho$ (left axis) and $d\rho/dT$ (right axis), over a narrow temperature range corresponding to the ordering temperatures. Dashed lines indicate feature in $d\rho/dT$ near $T_{\text{N}}$ and $T_{\text{SR}}$, marking the transition temperatures. These transition are consistent with those observed in $C_P$, and $\frac{d(\chi T)}{dT}$ measurements. RRR of approximately $2$ is observed.

The field-dependent magnetization (Fig. \ref{fig:Tb_based_A=000020}$d$) for \( H\parallel a \) initially shows a gradual increase with the applied magnetic field. Upon further increase in field strength, two distinct changes in slope are observed in the magnetization curve. The first transition, which occurs near 15 kOe where there is a increase in slope followed by a sharper increase in slope near 25 kOe. Beyond \( 35 \) kOe, the magnetization shows minimal variation, approaching a nearly constant value of approximately \( 8.62(22)\mu_B/\mathrm{Tb} \), which is fairly consistent with the theoretical saturation moment of Tb\(^{3+}\) (\(9~\mu_B\)) within uncertainty. When the field is reduced, the lower transition of these two meta-magnetic transitions disappears, leaving behind a small hysteresis loop. For $H \perp a$ the magnetization is small and roughly linear in H.  There is a small hysteresis feature and weak non-linearity that is most likely associated with small misalignment of the applied field.
 

In figure \ref{fig:Tb_based_A=000020}$e$, the inverse susceptibilities, $\chi_{\parallel}^{-1}$ and $\chi_{\perp}^{-1}$, extracted from the data in figure \ref{fig:Tb_based_A=000020}$b$, are presented. Curie-Weiss equation was used to fit the $\chi_{||}^{-1}$ and $\chi_{\perp}^{-1}$ data in order to obtain $\theta_{||} \approx 21$ K and $\theta_{\perp} \approx - 78$ K.  $\chi_{\text{powder}}$ and its inverse, $\chi_{\text{powder}}^{-1}$, are shown in figure \ref{fig:Tb_based_A=000020}$f$ for an applied magnetic field of 1 kOe and $\mu_{\text{eff}}$ = $9.97(18)$ $\mu_{\text{B}}/\text{Tb}$ was extracted by fitting the experimental data to the Curie-Weiss law, as shown by the solid line. This value closely matches the theoretical value for Tb$^{3+}$ within uncertainty, which is  $9.72$ $\mu_{\text{B}}$.

\subsubsection{$\text{Dy}_{2}\text{Co}_{6}\text{Al}_{20-\delta}$}

\begin{figure*}
\centering{}\includegraphics[width=0.80\paperwidth]{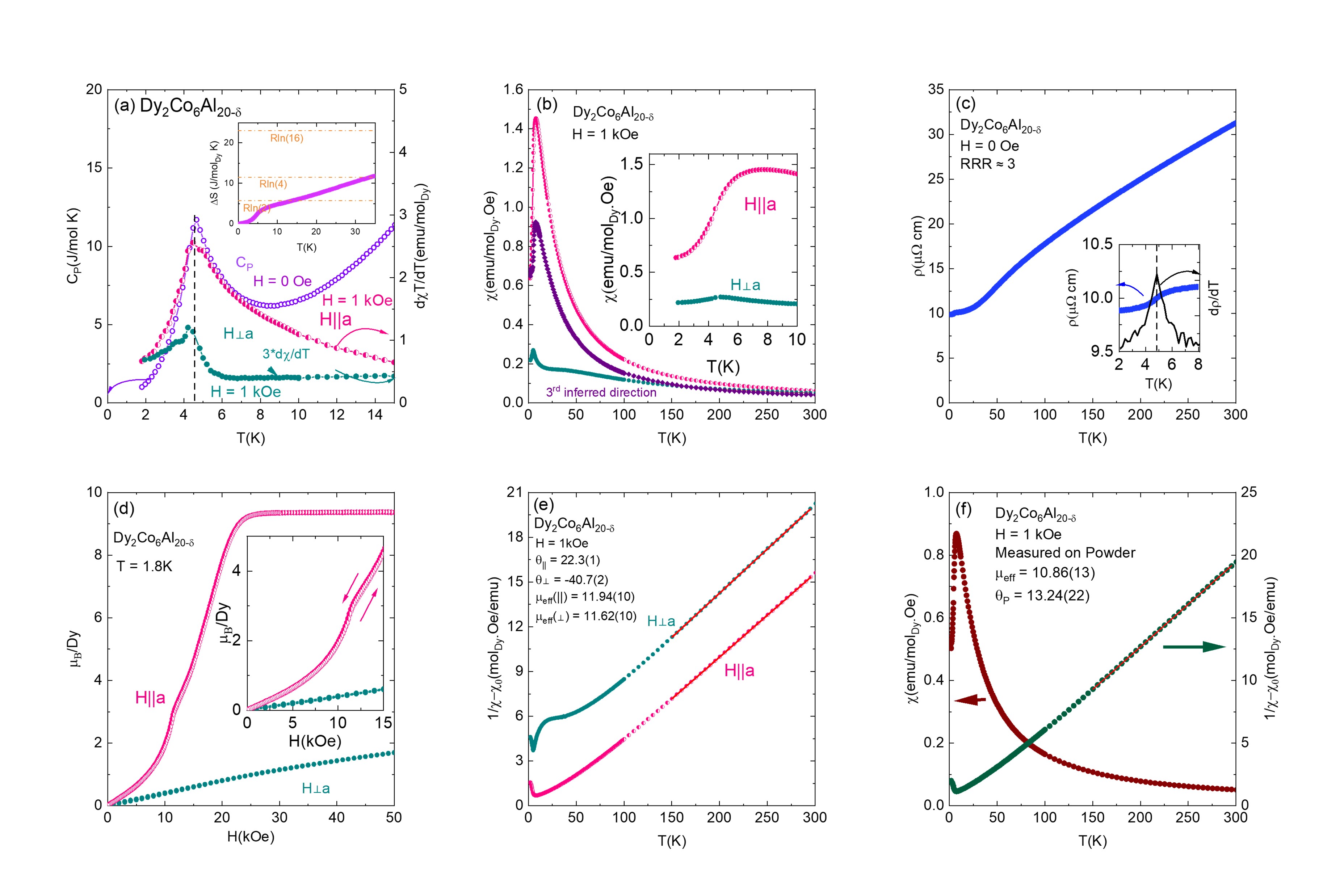} \caption{\footnotesize Physical properties of {Dy$_{2}$Co$_{6}$Al$_{20-\delta}$ single crystal.
$(a)$ $C_{p}$ (left axis), and $d(\chi T)/dT$ (right-axis) obtained for $H\parallel a$ ( half open symbols ) and $H\perp a$ ( solid symbols) with an applied field of $H=1$ kOe, inset: magnetic entropy(see text for detail). $(b)$ $\chi_{||}$ and $\chi_{\perp}$ showing the a AFM transition at low-$T$, and curie-Weiss like behavior at high temperature range as expected and inset showing close view  near transition temperature. $(c)$ Temperature dependent electrical resistivity measurements (I $||a$). In the inset, we present $\rho$ and $d\rho/dT$ in the near the AFM transitions. $(d)$ Isothermal magnetization measurements at $T=1.8$ K for applied field up to $50$ kOe for $H\parallel a$ ( half open symbols ) and $H\perp a$ ( solid symbols). $(e)$ Inverse susceptibility data was obtained for $H\parallel a$ and $H\perp a$ and used to extract parameters $\theta_{\text{||}}$ and $\theta_{\perp}$. $(f)$ $\chi_{\text{powder}}$ (left-axis and $(\chi_{\text{powder}}-\chi_{0})^{-1}$ (right-axis) data was used to extract $\mu_{\text{eff}}$}, and the solid lines in figure $(e)$ and $(f)$ show the Curie-Weiss fit used to estimate $\mu_{\text{eff}}$ and $\theta$.}
\label{fig:Dy_based_A=000020}
\end{figure*}

Figure \ref{fig:Dy_based_A=000020} summarizes the data we collected on $\text{Dy}_{2}\text{Co}_{6}\text{Al}_{20-\delta}$, the $C_p$ is plotted as a function of $T$ (left axis in figure \ref{fig:Dy_based_A=000020}$a$). The data show a pronounced peak at \( T = 5.0(4) \) K, indicative of a phase transition, followed by a monotonic decrease after the peak. In the inset of figure \ref{fig:Dy_based_A=000020}$a$, we present an estimate of the magnetic entropy, where we recover entropy approximately R$\ln 2$ around 13 K, which suggests doublet ground state. Above this temperature, the entropy continues to increase, similar to the behavior observed in Tb-based compound, consistent with higher-lying crystal electric field levels.

The temperature dependent magnetic susceptibilities curves for $\chi_{\parallel}$ and $\chi_{\perp}$ exhibit the characteristic of AFM ordering, as indicated by the observed peak-like behavior. The magnetic response of this compound exhibits pronounced anisotropy, as evidenced by the significantly larger magnetic susceptibility for the parallel direction compared to the perpendicular direction at low temperatures. At high temperatures, the behavior is Curie-Weiss-like in both $\chi_{\parallel}$ and $\chi_{\perp}$. The inset of figure \ref{fig:Dy_based_A=000020}$a$ provides a magnified view of the detailed behavior of the system at low temperatures. In figure \ref{fig:Dy_based_A=000020}$a$ (right axis), we plot the $\frac{d(\chi T)}{dT}$. Peak is observed around $ 5$ K for both directions  ($\chi_{\parallel}$ and $\chi_{\perp}$ ). This value is consistent, within the uncertainty, with the Néel temperature derived from $C_p$.

The resistivity of the Dy$_{2}$Co$_{6}$Al$_{20-\delta}$ single crystal (Fig. \ref{fig:Dy_based_A=000020}$c$), measured from $300$ K to $1.8$ K, exhibits a decrease upon cooling, similar to other compounds in the series. In particular, a distinct change in the slope of the resistivity curve is observed at low temperatures, indicating a possible transition. The inset highlights $\rho(T)$ (left axis) and its derivative, $d\rho/dT$ (right axis), over a temperature range of the ordering transition. Dashed lines identify transition in $d\rho/dT$ near $T_{\text{N}}$, marking the transition temperature, and the values are aligned with those obtained from $C_P$ and $\frac{d(\chi T)}{dT}$. RRR is determined to be approximately $3$.

Figure \ref{fig:Dy_based_A=000020}$d$ presents the isothermal magnetization measurements performed at $T = 1.8$ K for $H \parallel a$ and $H \perp a$, in the range $0 \leq H \leq 50$ kOe. As the applied magnetic field increases, two distinct transitions are observed in the magnetization. The initial magnetic transition is observed near \( H = 10 \, \mathrm{kOe} \), which may be because of spin flop transition similar like Tb case. A more prominent transition occurs around \( H = 25 \, \mathrm{kOe} \), characteristic of a metamagnetic transition. Beyond this field, the magnetization saturates gradually, reaching a value of \( 9.36(21) \, \mu_B \) /\text{Dy} , which is very close to the theoretical saturated moment of Dy$^{3+}$ (\(10.65 \, \mu_B\)).  The inset of figure \ref{fig:Dy_based_A=000020}$(d)$ provides a detailed view of the low-field region, highlighting data collected during both increasing and decreasing field to confirm the absence of hysteresis. The \( M(H) \) curves indicate that the magnetic moments predominantly align along the \( a \)-axis, as demonstrated by the occurrence of a metamagnetic transition in measurements taken with the magnetic field applied along the \( a \)-axis.

In figure \ref{fig:Dy_based_A=000020}$e$, the inverse susceptibilities, $\chi_{\parallel}^{-1}$ and $\chi_{\perp}^{-1}$ are shown. Furthermore, $\chi_{\text{powder}}$ and  $(\chi_{\text{powder}}-\chi_{0})^{-1}$ are presented in figure \ref{fig:Dy_based_A=000020}$f$, all obtained with  $H=1$ kOe. $\chi_{\parallel}$, $\chi_{\perp}$, and $\chi_{\text{powder}}$ exhibit Curie-Weiss (CW) behavior at high $T$. Using equation \ref{eqn:curie_R} (fitting has shown with solid line), $\mu_{\text{eff}}$ and Curie-Weiss temperature was extracted. Data from figure \ref{fig:Dy_based_A=000020}$e$ used to obtain \(\theta_{||} \approx 22\) K and \(\theta_{\perp} \approx - 41\) K.  From the data in figure \ref{fig:Dy_based_A=000020}$f$, we determine $\mu_{\text{eff}} = 10.86(13)$ $\mu_{\text{B}}$/Dy, which is in close to the theoretical value for Dy$^{3+}$, $10.65$ $\mu_{\text{B}}$.

\subsubsection{$\text{Ho}_{2}\text{Co}_{6}\text{Al}_{20-\delta}$}

\begin{figure*}
\centering{}\includegraphics[width=0.80\paperwidth]{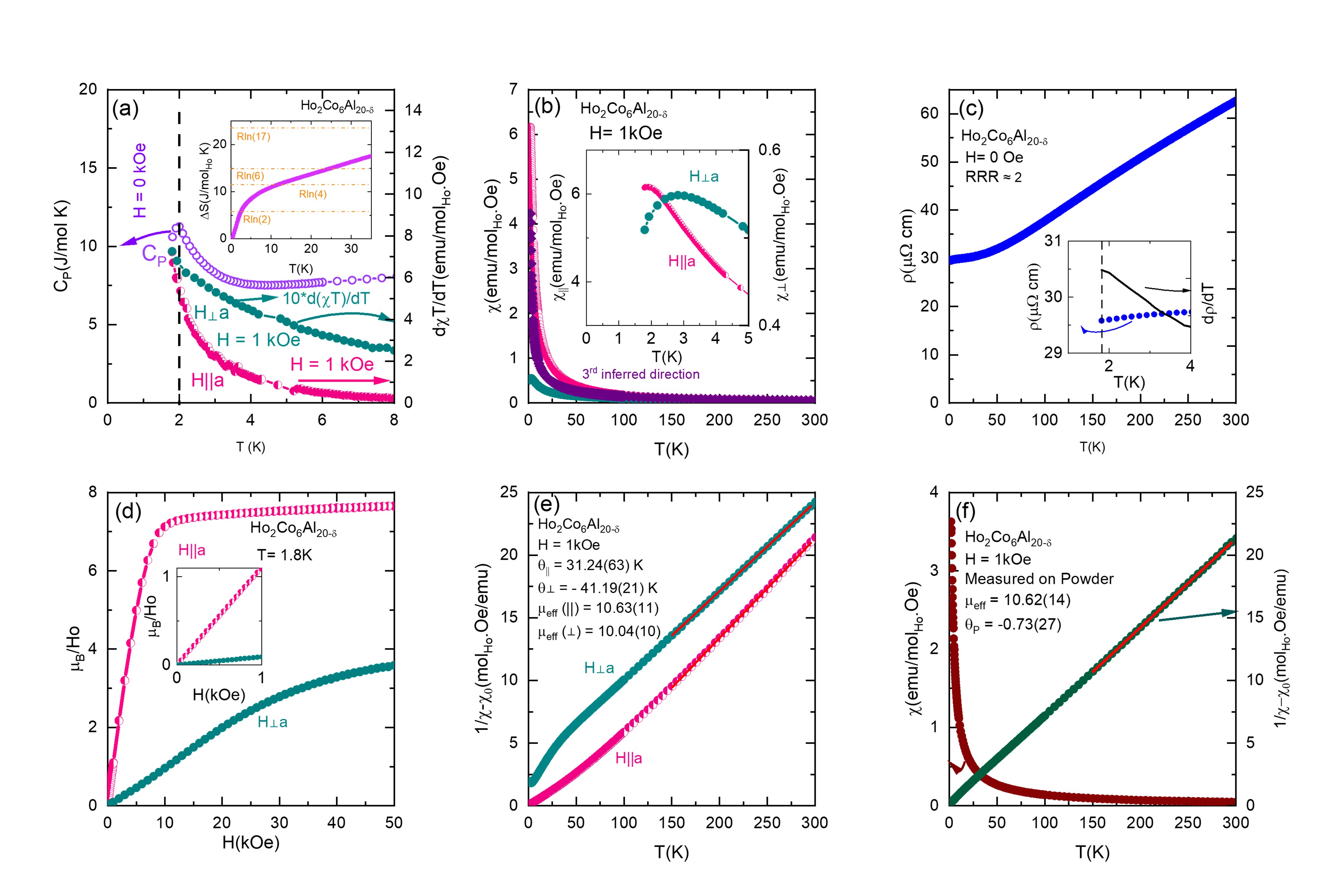}
\caption {\footnotesize{Physical properties of Ho$_{2}$Co$_{6}$Al$_{20-\delta}$ single crystal .
$(a)$ $C_{p}$ (left axis) and $d(T\chi)/dT$ (right-axis) obtained for $H\parallel a$ ( half open symbols ) and $H\perp a$ ( solid symbols) with an applied field of $H=1$ kOe. For \( H = 1 \, \text{kOe} \), the data does not exhibit a well-defined transition till \( T = 1.8 \, \text{K} \), inset: magnetic entropy(see detail in text). $(b)$ $\chi_{||}$ and $\chi_{\perp}$ was measured at $1$ kOe. The inset highlights the low-temperature region, where a clear downturn is observed, consistent with an antiferromagnetic transition. $(c)$Temperature dependent electrical resistivity measurements (I $||a$). In the inset, we present $\rho$ and $d\rho/dT$ near the AFM transitions. The AFM transition can be observed as a peak in the temperature derivative about $T_{N}$. $(d)$ Isothermal magnetization measurements $(T=1.8$ K) for applied field up to $5$ T for $H\parallel a$ ( half open symbols ) and $H\perp a$ ( solid symbols). $(e)$  $\chi^{-1}$ data obtained for $H\parallel a$ and $H\perp a$ to obtained parameters $\theta_{\text{||}}$ and $\theta_{\perp}$. $(f)$ $\chi_{\text{powder}}$ (left-axis) and $(\chi_{\text{powder}}-\chi_{0})^{-1}$ (right-axis) data was used to extract $\mu_{\text{eff}}$, and the solid lines in figure $(e)$ and $(f)$ show the Curie-Weiss fit used to estimate $\mu_{\text{eff}}$ and $\theta$.
 \protect\label{fig:Ho_based_A=000020}}}
\end{figure*}

\begin{figure}
\centering{}\includegraphics[width=0.4\paperwidth]{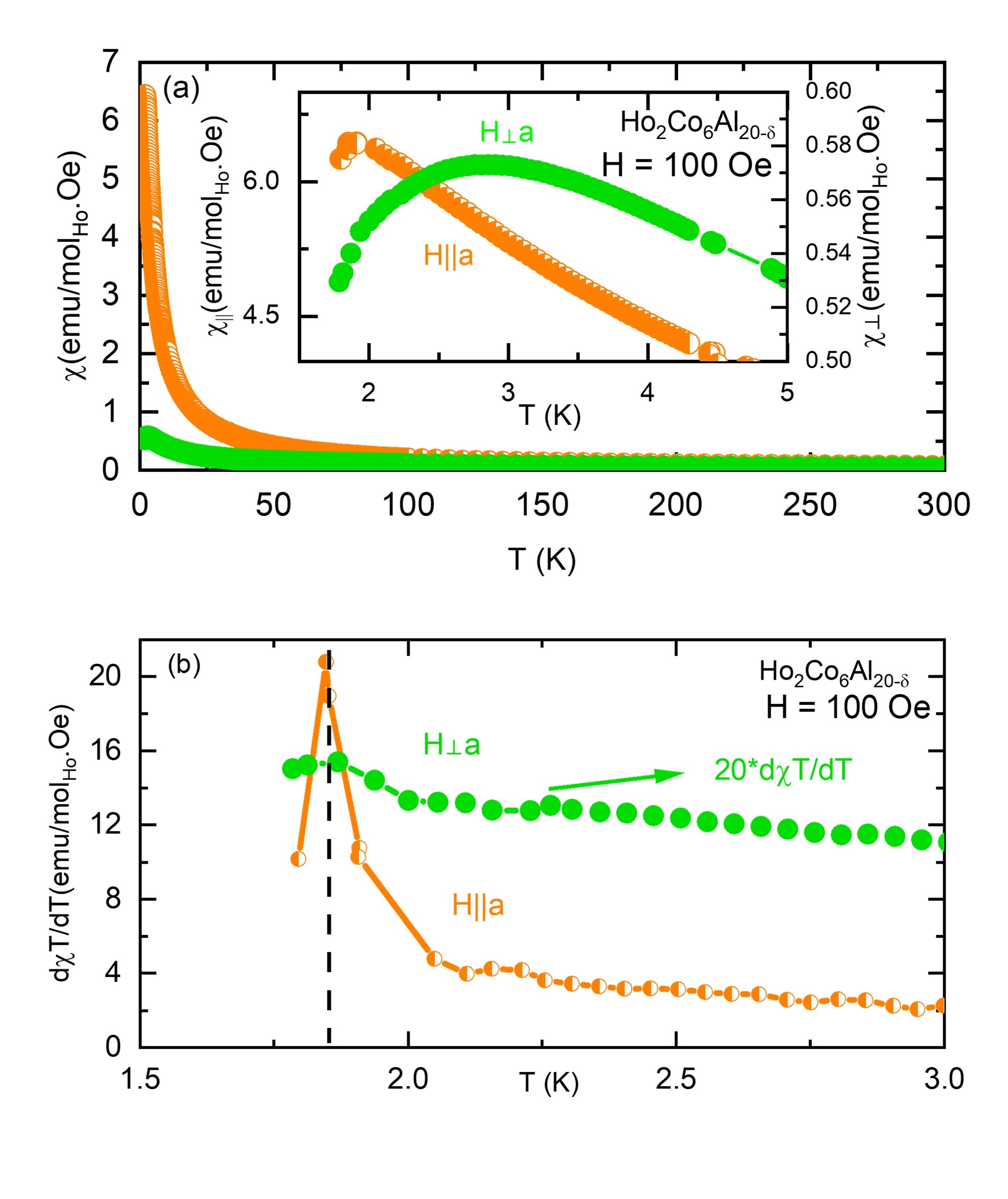}
\caption{\footnotesize{Temperature dependent magnetic data for Ho$_{2}$Co$_{6}$Al$_{20-\delta}$ measured at $100$ Oe. $(a)$ $\chi_{||}$ and $\chi_{\perp}$ ($H=100$ Oe) showing the a AFM transition at low-$T$ for both directions($\chi_{||}$ and $\chi_{\perp}$ ).$(b)$ Derivative of susceptibility, where we can clearly see the transition temperature around $1.8$ K(dashed line). \protect\label{fig:Ho_based_B}}}
\end{figure}

Figure \ref{fig:Ho_based_A=000020} summarizes the data we collected on Ho$_{2}$Co$_{6}$Al$_{20}$ . The  $ C_p (T)$ data exhibit a peak around $ 2.1(1)$ K as shown in figure \ref{fig:Ho_based_A=000020}$a$ (right axis).  Following the peak, $C_p$ decreases slightly upon cooling to 1.8 K. In the inest of figure \ref{fig:Ho_based_A=000020}$a$, we present the estimated magnetic entropy for Ho$_{2}$Co$_{6}$Al$_{20-\delta}$. We can see that the entropy for this compound does not recover its expected full entropy until 30 K, suggesting the presence of other Schottky-type contributions at higher temperatures. However, the inferred entropy for this compound is made uncertain due to the transition occurring near the lower measurement limit ($\approx 1.8 K$) and the specific heat being extrapolated to 0 K  for the entropy evaluation.

Temperature-dependent susceptibility curves at $1$ kOe exhibit a local maximum followed by a downturn near $1.8$ K for both orientations, as we can see in figure \ref{fig:Ho_based_A=000020}$b$. The compound exhibits pronounced magnetic anisotropy, as evidenced by the magnetic susceptibility being significantly higher for $H\parallel a$ compared to $H \perp a$, as we can see in the figure that we have a sharp increase at low temperature for $\chi_\parallel$ compared to $\chi_\perp$. To highlight subtle features of the system at low temperatures, a magnified view is provided in the inset. No transitions are observed in the $\frac{d(\chi T)}{dT}$ data up to $1.8$ K for either direction ( Fig. \ref{fig:Ho_based_A=000020}$a$, right axis). This is likely due to the fact that the anomaly in $\chi(T)$ occurs at the limit of the measured temperature range, making it difficult to resolve in the derivative data.

To explore further, susceptibility measurements were performed at a reduced magnetic field of $100$ Oe, revealing a peak-like feature (Fig. \ref{fig:Ho_based_B}). Subsequently, the derivative of $\chi$T was analyzed to determine the transition temperatures, which was clearly T$ \approx 1.8$ K for both $\chi_{\parallel}$ and $\chi_{\perp}$. These transition temperatures are consistent with the transition temperature obtained from the $C_p$ measurements.

In figure \ref{fig:Ho_based_A=000020}$c$, $\rho(T)$ is shown for an applied current ($I$) parallel to the crystallographic $a$-axis in the absence of an external magnetic field. The inset presents $\rho(T)$ (left axis) alongside its temperature derivative, $d\rho/dT$ (right axis), focusing on the region near the magnetic ordering transition. Dashed lines indicate the transition temperature, that is consistent with $T_{\text{N}}$ observed from  $C_P$ and $\frac{d(\chi T)}{dT}$. There is no clear feature in the $d\rho/dT$ data, but given the low RRR = 2 for this sample, combined with the low $T_N$ compared to our base temperature, it is not surprising that we cannot resolve a transition feature in $\rho(T)$.

Figure \ref{fig:Ho_based_A=000020}$d$ shows the isothermal magnetization \( M(H) \) measured at \( T = 1.8 \, \text{K} \) for applied magnetic fields aligned parallel and perpendicular to \( a \)-axis, over the range \( 0 \leq H \leq 50 \, \text{kOe} \). For \( H \parallel a\), the magnetization increases rapidly with the applied field up to approximately \( 10 \, \text{kOe} \), after which it's start saturating slowly up to \( 50 \, \text{kOe} \) and reaching to $7.63(22) \mu_B$/Ho, which is less than theoretical value for Ho$^{3+}$ ($10.61 \mu_B$). In contrast, for \( H \perp a \), the magnetization exhibits a rapid increase up to around \( 20 \, \text{kOe} \), followed by a slower growth at higher fields, as reflected in the reduced slope of the \( M(H) \) curve. Notably, saturation is not achieved for \( H \perp a \) within the measured field range. The inset of figure \ref{fig:Ho_based_A=000020}$d$ highlights the low-field region and includes data recorded during both increasing and decreasing field sweeps, confirming the absence of hysteresis. The \( M(H) \) curves indicate a strong anisotropy.

In figure \ref{fig:Ho_based_A=000020}$e$, the inverse magnetic susceptibilities, $\chi_{\parallel}^{-1}$ and $\chi_{\perp}^{-1}$, are plotted. Additionally, $\chi_{\text{powder}}$, and its inverse, $\chi_{\text{powder}}^{-1}$, are shown in figure \ref{fig:Ho_based_A=000020}$f$ for an applied magnetic field of 1 kOe. $\chi_{\parallel}$, $\chi_{\perp}$, and $\chi_{\text{powder}}$ follow the Curie-Weiss law at high temperature as expected. In figure \ref{fig:Ho_based_A=000020}$e$, a  Curie-Weiss fit (solid line) is performed to $\chi_{\parallel}^{-1}$ and $\chi_{\perp}^{-1}$ above 150 K in order to determined the Curie-Weiss $\theta$ as \(\theta_{\parallel} \approx 31 \, \text{K}\) and \(\theta_{\perp} \approx - 41 \, \text{K}\). From figure \ref{fig:Ho_based_A=000020}$f$, the effective magnetic moment, $\mu_{\text{eff}}$, was determined by fitting the susceptibility data to Curie-Weiss fit, represented by the solid line in the region above $150$ K. The extracted value, $\mu_{\text{eff}} = 10.66(14)$ $\mu_{\text{B}}$/Ho, aligns closely with the theoretical value for Ho$^{3+}$, calculated to be $10.61$ $\mu_{\text{B}}$.

\subsubsection{$\text{Er}_{2}\text{Co}_{6}\text{Al}_{20-\delta}$}

\begin{figure*}
\centering \includegraphics[width=0.80\paperwidth]{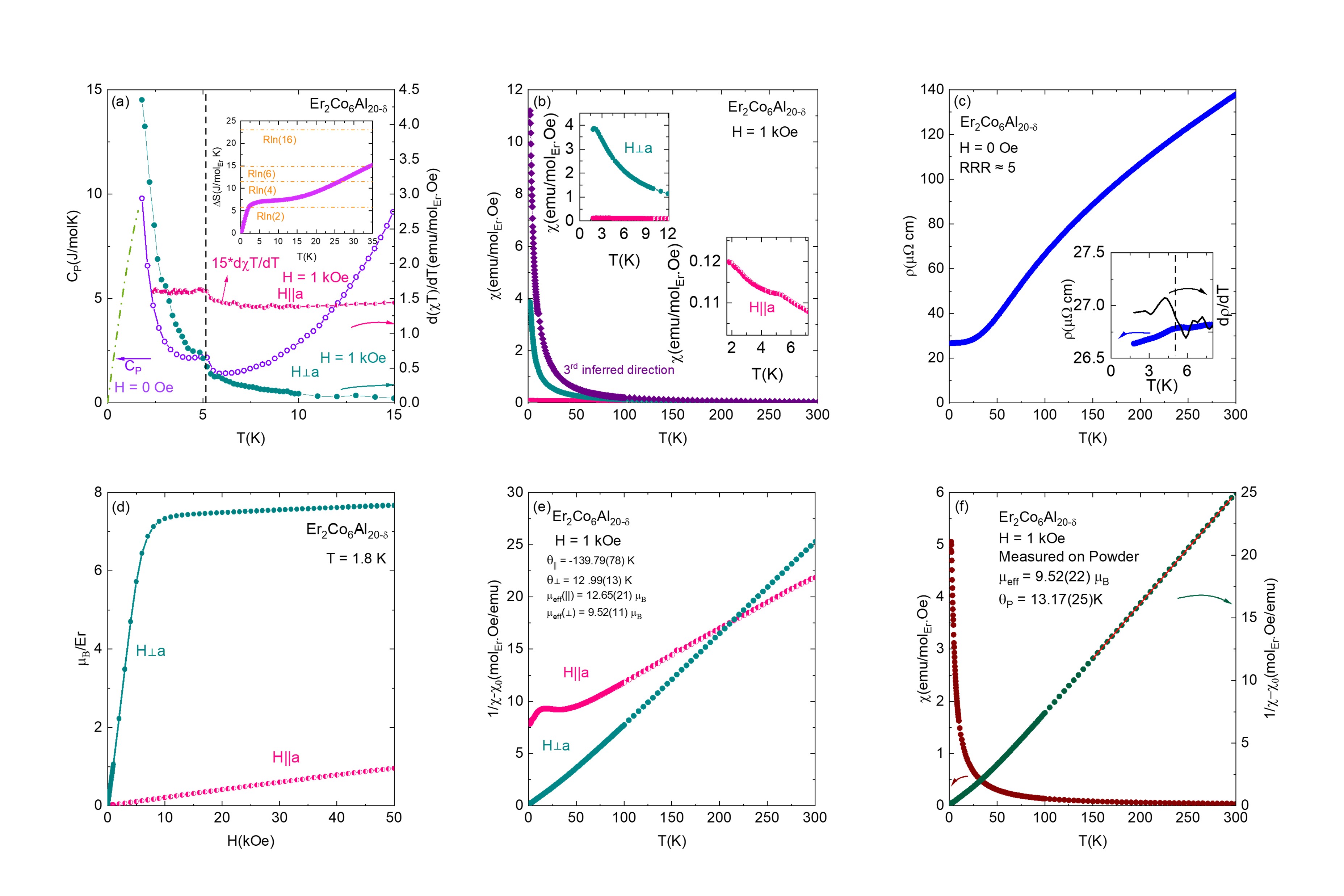} 
\caption{\footnotesize{ Physical properties of Er$_{2}$Co$_{6}$Al$_{20-\delta}$ single crystal. $(a)$ $C_{p}$ (left axis), where green dashed line showing extrapolation that we used to calculated entropy and $d(T\chi)/dT$ (right-axis) obtained for $H\parallel a$ ( half open symbols ) and $H\perp c$ ( solid symbols) with an applied field of $H=1$ kOe. The data shows a feature around  $T_N =5.4$ K  in both $C_{p}$  and $d(T\chi)/dT$ ($H\parallel a $), inset: magnetic entropy inferred from $C_P$ data(see details in text). $(b)$ Anisotropic susceptibility at H = $1$kOe for $\chi_{||}$ and $\chi_{\perp}$ is shown. The upper inset of this figure highlights the data for $H \perp a$, while the lower inset emphasizes the feature around 5 K for $H \parallel a$. $(c)$ Temperature dependent electrical resistivity measurements (I $||a$). $(d)$ Isothermal magnetization measurements $(T=1.8$ K) for applied field up to $50$ kOe for $H\parallel a$ ( half open symbols ) and $H\perp a$ (solid symbols).  $(e)$ Inverse susceptibility data obtained for $\chi_\parallel $ and $\chi_\perp $ and $\theta_{\text{||}}$ and $\theta_{\perp}$ parameters were obtained using this inverse susceptibility. $(f)$ $\chi_{\text{powder}}$ (left-axis) and $(\chi_{\text{powder}}-\chi_{0})^{-1}$ (right-axis) data used to extract $\mu_{\text{eff}}$. In figure (e) and (f) solid line correspond to the Curie-Weiss fits, from which the $\theta$ and the effective magnetic moment were extracted.
\label{fig:Er_based_A=000020}}}
\end{figure*}

Figure \ref{fig:Er_based_A=000020} summarizes the data we collected on Er$_{2}$Co$_{6}$Al$_{20-\delta}$. The $C_p (T)$ data (Fig. \ref{fig:Er_based_A=000020}$a$) exhibit  a peak-like feature near \(T_N = 5.4(2) \, \text{K}\)  and  a continuous increase upon further cooling below the transition, suggesting the possibility of spin-reorientation transition below $1.8$ K. To estimate the magnetic entropy of $\text{Er}_{2}\text{Co}_{6}\text{Al}_{20-\delta}$, a linear extrapolation of the specific heat data was performed from the lowest measured temperature 1.8 K down to $0$ K. While this provides a useful approximation, the resulting values represent an underestimation of the true entropy, as the simple linear extrapolation may not fully account for the complexity of the low-temperature magnetic contributions. The estimated entropy of this compound approaches $\approx$ R$ln2$, which suggest that ground state is doublet (shown in the inset of fig. \ref{fig:Er_based_A=000020}$a$). However, entropy continuous increases until 30 K, suggesting high-lying CEF levels. It must be noted that the calculated $S_{\text{mag}}$ values have a higher degree of uncertainty. 

 We have measured temperature dependent susceptibility for both direction($\chi_\parallel$ and $\chi_\perp$). Our susceptibility measurements exhibit anisotropic behavior at temperatures up to approximately 100 K, with $\chi_{\perp}$ being substantially greater than $\chi_{\parallel}$ in this temperature range (Fig. \ref{fig:Er_based_A=000020}$b$). This behavior contrasts with other compounds that we have measured so far, indicating a change in the sign of magnetic anisotropy when moving from Ho to Er. A subtle peak is observed around $T = 5$ K in $\chi_{\parallel}$, though it does not show a sharp peak. Below this temperature, the susceptibility exhibits a continuous increase toward lower temperatures. On the other hand, $\chi_{\perp}$ shows a slight downturn at low temperatures. At higher temperatures, the magnetic behavior in both the $\chi_{\parallel}$ and $\chi_{\perp}$ directions is in agreement with the Curie-Weiss law. The powder-averaged susceptibility much larger than the susceptibilities measured along the two crystallographic directions, implying that magnetic moment along another $\chi_{\perp}$ direction. In figure \ref{fig:Er_based_A=000020}$a$ (right axis), the derivative $\frac{d(\chi T)}{dT}$ is plotted for both directions. For $\chi_{\parallel}$, a subtle peak is observed near $\approx 5.2$ K, aligning with the peak detected in the $C_p$ measurements, suggesting that the system undergoes a magnetic phase transition at this temperature. Also in $\chi_{\perp}$, we have observed a kink around $\approx 5.2$ K that is consistent with our transition observed in $C_P$.

In figure \ref{fig:Er_based_A=000020}$c$, $\rho(T)$ is plotted for an applied current ($I$) aligned along the crystallographic $a$-axis, with no external magnetic field. The inset displays $\rho(T)$ (left axis) and its temperature derivative, $d\rho/dT$ (right axis), highlighting the region near magnetic ordering transition. Dashed lines denote a characteristic feature in $d\rho/dT$, identifying the Néel temperature ($T_{\text{N}}$). This transition aligns with those observed in $C_P$ and $\frac{d(\chi T)}{dT}$.

Figure \ref{fig:Er_based_A=000020}$d$ presents the isothermal magnetization \( M(H) \) curves measured at \( T = 1.8 \, \text{K} \) for $ H \parallel a $ and $H \perp a$, over the range \( 0 \leq H \leq 50 \, \text{kOe} \). For \( H \perp a \), \( M(H) \) exhibits a rapid increase with field strength up to approximately \( 10 \, \text{kOe} \) and then increases very slowly to \( 50 \, \text{kOe} \), reaching $7.68 \mu_B/ Er$, which is smaller than theoretically $\mu_{sat}$ ( $9 \mu_B$) for Er$^{+3}$. In contrast, for \( H \parallel a \), the magnetization increases gradually throughout the field range, showing a linear behavior without apparent saturation. These observations highlight significant magnetic anisotropy, with the magnetic moments preferentially aligning along the direction perpendicular to the \( a \)-axis, as indicated by the larger magnetization values for \( H \perp a \) compared to \( H \parallel a \).

 $\chi_{\parallel}^{-1}$ and $\chi_{\perp}^{-1}$(Fig. \ref{fig:Er_based_A=000020}$e$) have been plotted to extract Curie-Weiss temperature in order to understand magnetic anisotropy.  $\chi_{\text{powder}}$, and  $({\chi{_\text{powder} - \chi_0}})^{-1}$, are shown in figure \ref{fig:Er_based_A=000020}$(f)$ under an applied magnetic field of $1$ kOe. The susceptibilities $\chi_{\parallel}$, $\chi_{\perp}$, and $\chi_{\text{powder}}$ behave like the Curie-Weiss law at high temperature as expected. From Curie-Weiss fit (solid line, Fig. \ref{fig:Er_based_A=000020}$e$), the Curie-Weiss temperatures were extracted as \(\theta_{\parallel} \approx - 140 \, \text{K}\) and \(\theta_{\perp} \approx 13 \, \text{K}\), reflecting a combination of CEF-splitting effects as well as the possible anisotropic nature of the magnetic interactions. From figure \ref{fig:Er_based_A=000020}$f$, the effective magnetic moment, $\mu_{\text{eff}}$, was extracted by fitting the magnetic susceptibility data to the Curie-Weiss law, depicted as the solid line in the temperature range above $150$ K. The obtained value, $\mu_{\text{eff}} = 9.52(22)$ $\mu_{\text{B}}$/Er, is in close agreement with the theoretical value for Er$^{3+}$ ($S = \frac{3}{2}$), calculated as $9.58$ $\mu_{\text{B}}$.

\subsubsection{$\text{Tm}_{2}\text{Co}_{6}\text{Al}_{20-\delta}$}

\begin{figure*}
\centering \includegraphics[width=0.80\paperwidth]{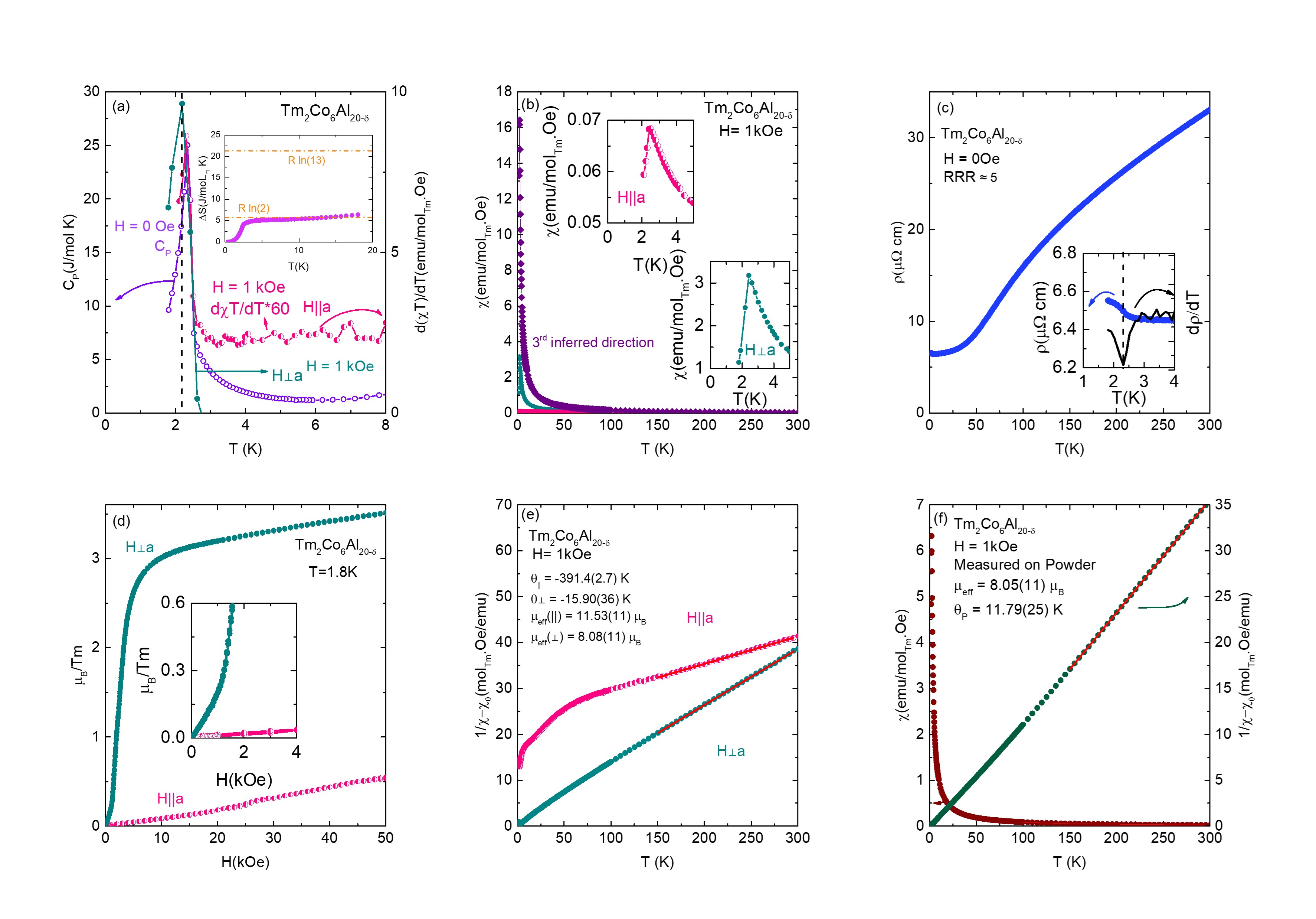} 
\caption{\footnotesize{ Physical Properties of Tm$_{2}$Co$_{6}$Al$_{20-\delta}$ single crystal. $(a)$ $C_{p}$ (left axis) and $d(T\chi)/dT$ (right-axis) obtained for $H\parallel a$ ( half open symbols ) and $H\perp a$ ( solid symbols) with an applied field of $H=1$ kOe. The data suggest an AFM transition taking place at with $T_{\text{N}}=2.4$ K. (b)Temperature dependent susceptibility was measured at $1$ kOe field for both directions. $(c)$ $\rho$ measurements (I $||a$) as a function of $T$. A $RRR$ of about $5$ is observed. In the inset, we present $\rho$ and $d\rho/dT$ in the range of the transition. $(d)$ Isothermal magnetization measurements at $T=1.8$ K for applied field up to $50$ kOe for $H\parallel a$ ( half open symbols ) and $H\perp a$ (solid symbols). $(e)$ $\chi^{-1}$ data obtained for $\chi_{\parallel} $ and $\chi_{\perp} $. Parameters $\theta_{\text{||}}$ and $\theta_{\perp}$ were obtained using Curie-Weiss fitting to this data. $(f)$ $\chi_{\text{powder}}$ (left-axis) and $(\chi_{\text{powder}}-\chi_{0})^{-1}$ (right-axis) data was used to extract $\mu_{\text{eff}}$. The solid lines in figure (e) and (f) represent Curie-Weiss fits applied to the high-temperature susceptibility data, from which $\theta_{CW}$ and $\mu_{\text{eff}}$ were determined.}
\label{fig:Tm_based_A=000020}}
\end{figure*}

Figure \ref{fig:Tm_based_A=000020} summarizes the data we collected on Tm$_{2}$Co$_{6}$Al$_{20-\delta}$.  We have observed a sharp peak for the $C_p$ data (Fig. \ref{fig:Tm_based_A=000020}$a$, left axis) at the temperature, $T_N = 2.4(1)$ K, and we can see a monotonically decrease after transition. In inset of figure \ref{fig:Tm_based_A=000020}$a$, we have shown estimated magnetic entropy for Tm${_2}$Co${_6}$Al${_{20-\delta}}$ upto around $17$ K that appears close to R$ln2$, which indicates that the ground state most likely is a pseudo-doublet.

The $\chi_{\parallel}$ and $\chi_{\perp}$ curves reveal AFM ordering, evidenced by the peak-like shape in $\chi(T)$ in figure \ref{fig:Tm_based_A=000020}$b$. The inset of figure \ref{fig:Tm_based_A=000020}$b$ is showing a detailed view behavior of the system at low temperatures for $\chi_{\parallel}$ (upper inset) and $\chi_{\perp}$(lower inset). Similar to the Er-based compound, the Tm-based compound exhibits a large powder-averaged susceptibility, indicating that the magnetic moments are likely oriented along a different crystallographic direction. A peak is observed in both direction $\chi_{\parallel}$ and $\chi_{\perp}$ in $\frac{d(\chi T)}{dT}$ data ( Fig. \ref{fig:Tm_based_A=000020}$a$, right axis) at characteristic temperatures $T_N \approx 2.4$ K . These peaks closely correspond, within the uncertainty range, to the peak observed in the $C_p$ data.

The resistivity data (Fig. \ref{fig:Tm_based_A=000020}$c$) exhibit an upturn at low temperatures, potentially indicative of the opening of a superzone gap. The inset illustrates $\rho(T)$ (left axis) and its temperature derivative $d\rho/dT$ (right axis) over a temperature range near the magnetic ordering transition. The dashed line shows a local minima in $d\rho/dT$ corresponding to the transition temperature that was observed in $C_P$ and the derivative $\frac{d(\chi T)}{dT}$ measurements. RRR is calculated to be approximately $5$.


The isothermal magnetization measurements were collected at $T = 1.8$ K for both $H \parallel a$ and $H \perp a$ over the range $0 \leq H \leq 50$ kOe (fig. \ref{fig:Tm_based_A=000020}$d$). For $H \perp a$, the magnetization initially exhibits a linear response up to $H = 1$ kOe, followed by a sharp jump, characteristic of a metamagnetic transition. Above \( H \approx 10 \, \text{kOe} \), the magnetization gradually increases, reaching 3.46(25)$\mu_B$/Tm$^{3+}$ at \( 50 \, \text{kOe} \), which remains well below the theoretical saturation moment of $7.56 \mu_B$ expected for  $\text{Tm}^{3+}$. This is consistent with our inference that the second, $H \perp a$ direction has a much larger $\chi(T)$ behavior, due most likely to large CEF-induced anisotropy. The inset in figure \ref{fig:Tm_based_A=000020}$d$ focuses on the low-field regime, providing detailed data for both increasing and decreasing fields, confirming the absence of hysteresis and emphasizing the metamagnetic transition. The \( M(H) \) curves suggest that the magnetic moments predominantly align perpendicular to the \( a \)-axis, as evidenced by the observed metamagnetic transition for $H \perp a$.

In figure \ref{fig:Tm_based_A=000020}$e$, the inverse susceptibilities, $\chi_{\parallel}^{-1}$ and $\chi_{\perp}^{-1}$, derived from the data in figure \ref{fig:Tm_based_A=000020}$b$, are presented. Additionally, $\chi_{\text{powder}}$, and its inverse, $\chi_{\text{powder}}^{-1}$, are shown in figure \ref{fig:Tm_based_A=000020}$f$ for an applied magnetic field of $1$ kOe. $\chi_{\parallel}$, $\chi_{\perp}$, and $\chi_{\text{powder}}$ follow Curie-Weiss (CW) behavior at high temperatures. $\theta_{\parallel}$ and $\theta_{\perp}$ were calculated from  $\chi_{\parallel}^{-1}$ and $\chi_{\perp}^{-1}$ data set. Meanwhile, $\chi_{\text{powder}}$ was used to determine $\mu_{\text{eff}}$. A  Curie-Weiss fit (solid lines) is performed to $\chi_{\parallel}^{-1}$ and $\chi_{\perp}^{-1}$ data to get \(\theta_{\parallel} = - 391\) K and \(\theta_{\perp} = - 16\) K, reflecting the distinct nature of magnetic interactions along different direction. From figure \ref{fig:Tm_based_A=000020}$f$, the effective magnetic moment, $\mu_{\text{eff}}$, was obtained by fitting the data to the Curie-Weiss law, indicated by the solid line. The extracted value, $\mu_{\text{eff}} = 8.05(11)$ $\mu_{\text{B}}$/Tm, shows good agreement with the theoretical moment for Tm$^{3+}$ ($S = 1$), calculated as $7.6$ $\mu_{\text{B}}$.

\begin{table*}[t]
\centering
\caption{The table below presents the moment orientation direction, $\mu_{\text{eff}}$, $\mu_{\text{sat}}$ and RRR. Here, $T_N$ denotes the Néel temperature corresponding to the magnetic ordering, and $T_{SR}(K)$ indicates possible spin reorientation transition at a lower temperature. The transitions listed in the table are derived from $C_p$ data, while transitions identified from susceptibility measurements are discussed in detail in the main text for each compound. Notably, the transitions observed in both datasets exhibit excellent agreement within uncertainty.}

\label{tab:table}
\addtolength{\tabcolsep}{-2pt} 
\small
\begin{tabular*}{\textwidth}{@{\extracolsep{\fill}}lcccccccc}
\toprule
Compounds & \makecell{Easy \\ Orientation} & $T_N $ (K)& \makecell{$T_{SR}$ \\ (K)} & \makecell{$\mu_{eff}$ \\ ($\mu_B$)} & \makecell{$\mu_{sat}$ \\ ($\mu_B$)} & RRR & $\chi_0(emu/mol_R.Oe)$ \\ \hline
$\text{Y}_{2}\text{Co}_{6}\text{Al}_{20-\delta}$ & - & - & - & - & - & 2 & 3.7(14) $\times$ $10^{-4}$\\ \hline
$\text{Gd}_{2}\text{Co}_{6}\text{Al}_{20-\delta}$ & - & 9.7(3) & 8.2(3) & 8.16(21) & - & 2 & 2.0(7) $\times$ $10^{-3}$\\ \hline
$\text{Tb}_{2}\text{Co}_{6}\text{Al}_{20-\delta}$ & $\parallel a$-axis & 11.8(3) & $\approx 8$ & 9.97(18) & 8.62(22) & 2  & 9.4(4) $\times$ $10^{-3}$ \\ \hline
$\text{Dy}_{2}\text{Co}_{6}\text{Al}_{20-\delta}$ & $\parallel a$-axis & 5.0(4) & - & 10.86(13) & 9.36(21) & 3  & -4.4(44) $\times$ $10^{-4}$\\ \hline
$\text{Ho}_{2}\text{Co}_{6}\text{Al}_{20-\delta}$ & $\parallel a$-axis & 2.1(1) & - & 10.66(14) & 7.63(22) & 2 & -2.7(3) $\times$ $10^{-6}$\\ \hline
$\text{Er}_{2}\text{Co}_{6}\text{Al}_{20-\delta}$ & $\perp a$-axis & 5.4(2) & - & 9.52(22) & 7.64(35) & 5 & -1.7(18) $\times$ $10^{-4}$ \\ \hline
$\text{Tm}_{2}\text{Co}_{6}\text{Al}_{20-\delta}$ & $\perp a$-axis & 2.4(1) & - & 8.05(11) & 3.49(25) & 5 & -2.4(2) $\times$ $10^{-3}$ \\ \hline
 \bottomrule
 \end{tabular*}
 \end{table*} 
 
\section{Summary and Conclusions}

\begin{figure}
\centering \includegraphics[scale=0.35]{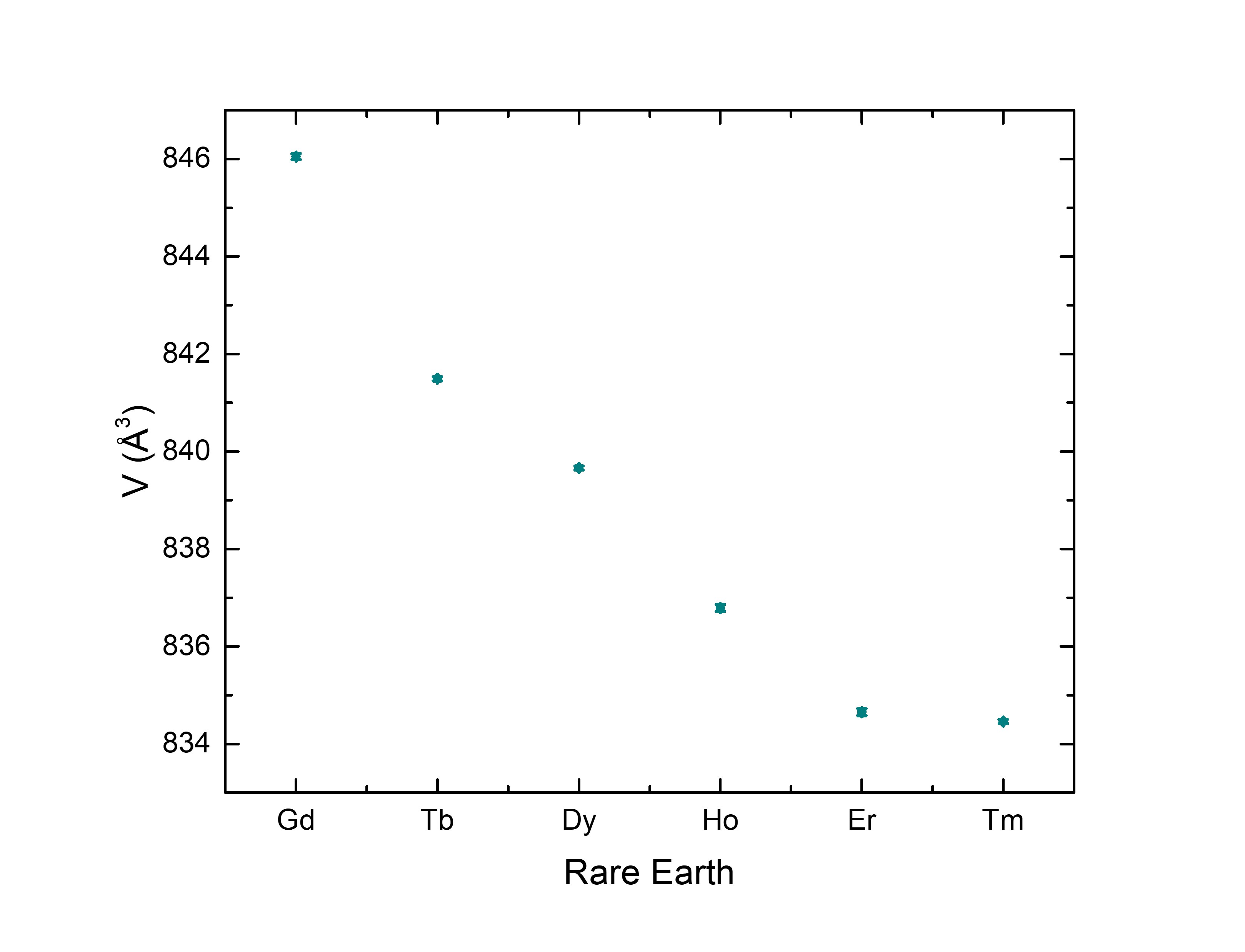}

\caption{\footnotesize{The unit cell volume of $R$$_{2}$Co$_{6}$Al$_{20-\delta}$ compounds, as a function of the rare earth element, was extracted from single crystal XRD.}}
\protect\label{fig:lattice_parameter}
\end{figure}

\begin{figure}
\centering{}\includegraphics[width=0.4\paperwidth]{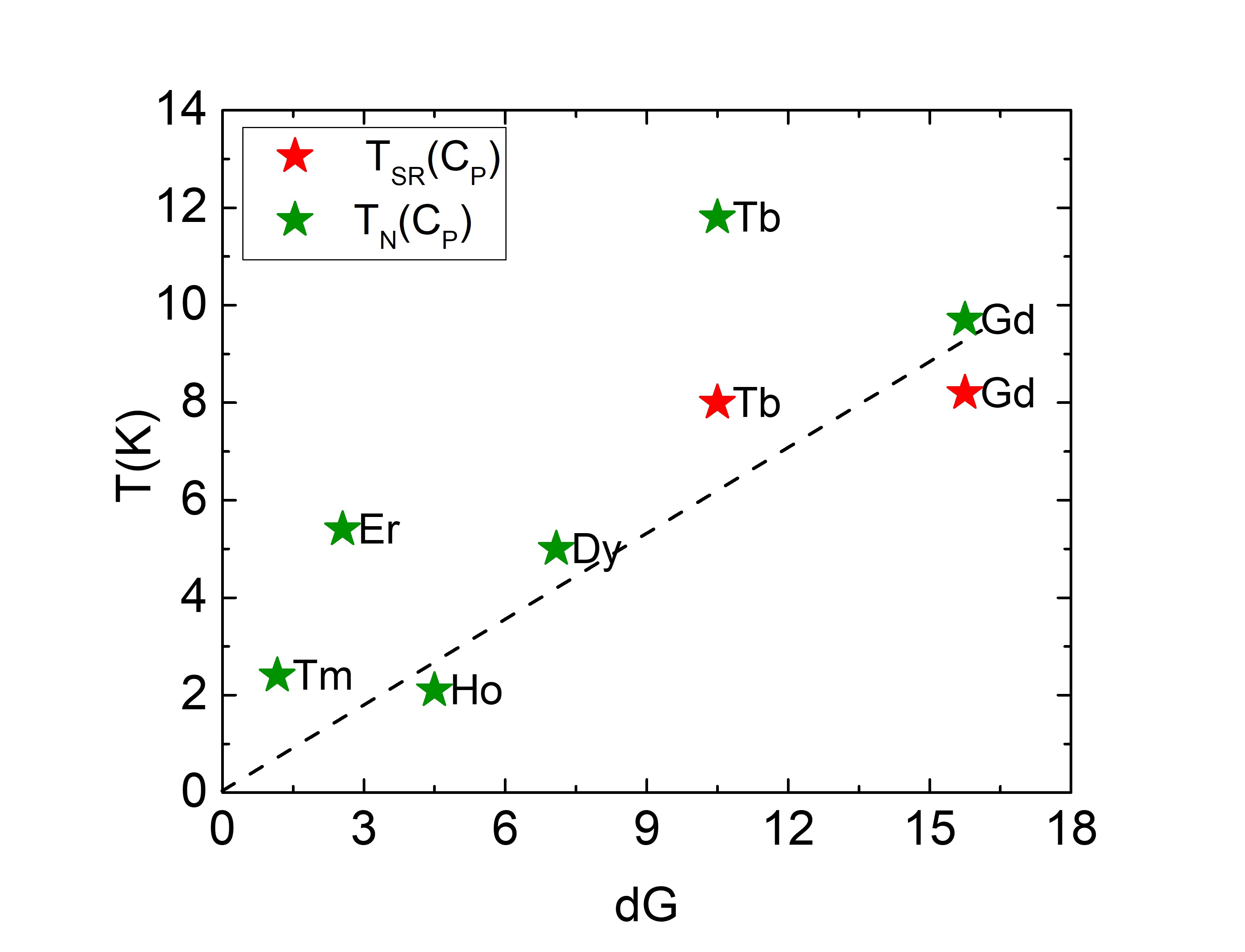}
\caption{\footnotesize{The plot of \(  dG \) scaling versus the transition temperatures, derived from \( C_p \) data, indicates that this system deviates from the expected \( \ dG \) scaling behavior.
\protect\label{fig:dG}}}
\label{fig:Ho2}
\end{figure}

The ${R_2}${Co$_6$}{Al$_{20-\delta}$} (where $R$ = Y, Gd-Tm) single crystals were initially discovered through arc melting \citep{R2Co6Al9poly}. However, its magnetic properties remained unexplored until now. For the first time, the single crystals of these materials have been grown using the solution growth technique. Comprehensive characterization experiments, including SCXRD, $C_p$, $\rho$ and $\chi$, were presented and discussed along the text. 

The single crystals have allowed for a more accurate characterization of the crystal structure. Whereas the family was initially suggested to adopt a monoclinic $C2/m$ structure with composition $R_{2}$Co$_{6}$Al$_{19}$, our SCXRD results convincingly show the structure is orthorhombic ($Imma$) and the composition better described as $R_{2}$Co$_{6}$Al$_{20-\delta}$. The SCXRD data is shown in Table \ref{tab:CrystalSturcture_Information}. Physical property measurements characterized all   ${R_2}${Co$_6$}{Al$_{20-\delta}$} compounds as low temperature AFM materials.  Table \ref{tab:table} brings a compilation of the data obtained from these measurements compounds.

In figure \ref{fig:lattice_parameter}, where the unit cell volume for each $R$-based material is presented as a functions of $R$ (data obtained from SCXRD data).  The trend in volume variation reflects the lanthanide contraction effect, where the unit cell volume decreases with increasing atomic number. Furthermore, Fig.~\ref{fig:dG} shows the Néel temperature $T_N$, determined from specific-heat measurements, plotted as a function of the de~Gennes scaling factor
\begin{equation}\label{eqn:dG}
dG = (g_J - 1)^2 J(J + 1),
\end{equation}
where $g_J$ is the Landé $g$-factor and $J$ is the total angular-momentum quantum number of the rare-earth ion. Within mean-field theory, the ordering temperature is expected to scale with the de~Gennes factor for systems dominated by purely RKKY exchange interactions, in the absence of strong crystal electric field (CEF) anisotropy or Kondo effects. Indeed, many rare-earth intermetallic compounds approximately follow de~Gennes scaling \citep{PhysRevB.80.104403,BUDKO199953}.

In contrast, $R_{2}$Co$_6$Al$_{20-\delta}$ compound breaks de~Gennes scaling, as evidenced by the pronounced shifts observed when moving from Gd to Tb and from Ho to Er. The pronounced deviation from linear de Gennes scaling suggests that magnetic ordering in this system is not governed solely by isotropic RKKY exchange, but is strongly influenced by crystal electric field anisotropy and exchange anisotropy. Similar departures from de Gennes behavior have been reported in the $R$Rh$_4$B$_4$ system, where the maximum ordering temperature occurs for DyRh$_4$B$_4$. It has been suggested that, in magnetically ordered systems, CEF effects can enhance $T_N$ in materials exhibiting strong axial anisotropy \citep{PhysRevB.29.6244,NOAKES198235}.

Consistent with this picture, our compounds also exhibit a crossover in magnetic anisotropy across the rare-earth series. Comparable anisotropy crossovers have been observed in several rare-earth intermetallic systems, including $R$AgSb$_2$ \citep{myers_systematic_1999} (tetragonal point symmetry at the rare-earth site), $R$Rh$_3$Si$_7$ \citep{PhysRevB.29.6244}(trigonal point symmetry), and $R$AgGe \citep{MOROSAN2004298} and $R$PtIn \citep{PhysRevB.72.014425}(orthorhombic point symmetry). For tetragonal point symmetry, as in $R$AgSb$_2$\citep{myers_systematic_1999}, changes in magnetic anisotropy can often be understood in terms of a sign change of the $B_2^0$ CEF parameter when moving from Ho to Er. However, in orthorhombic point symmetry, many CEF parameters are allowed, making the anisotropy behavior more complex. Consequently, a full CEF analysis is required to understand the observed anisotropy crossover.

Also, we have observed reduced entropy above ordering temperature in Gd based compound. The origin of the reduced magnetic entropy remains an open question and likely arises from several factors. It can be a combination of magnetic fluctuation, structure change, and quasi 1D structure.

In summary, we presented the successful single crystal growth and characterization of  $R_{2}$Co$_{6}$Al$_{20-\delta}$ compounds. All materials undergo an  AFM transition at low temperatures and (with the exception of the Gd-based materials) present strong CEF induced magnetic anisotropy. The Gd and Tb-based compound exhibits two magnetic transition transitions based on our data, where the lower-temperature transition is likely a spin-reorientation transition. To get a more detailed understanding of these transitions, neutron diffraction measurements are necessary.

\FloatBarrier

\begin{acknowledgments} This work was supported by the U.S. Department of Energy, Office of Basic Energy Science, Division of Materials Sciences and Engineering. The research was performed at the Ames Laboratory. Ames Laboratory
is operated for the U.S. Department of Energy by Iowa State University under Contract No. DE-AC02-07CH11358 Financial support from CNPq grant No. 444081/2024-0 is acknowledged by F.A.G and R.A.R.  
\end{acknowledgments}

\section*{Appendix A}

\begin{figure}
\centering \includegraphics[scale=0.30]{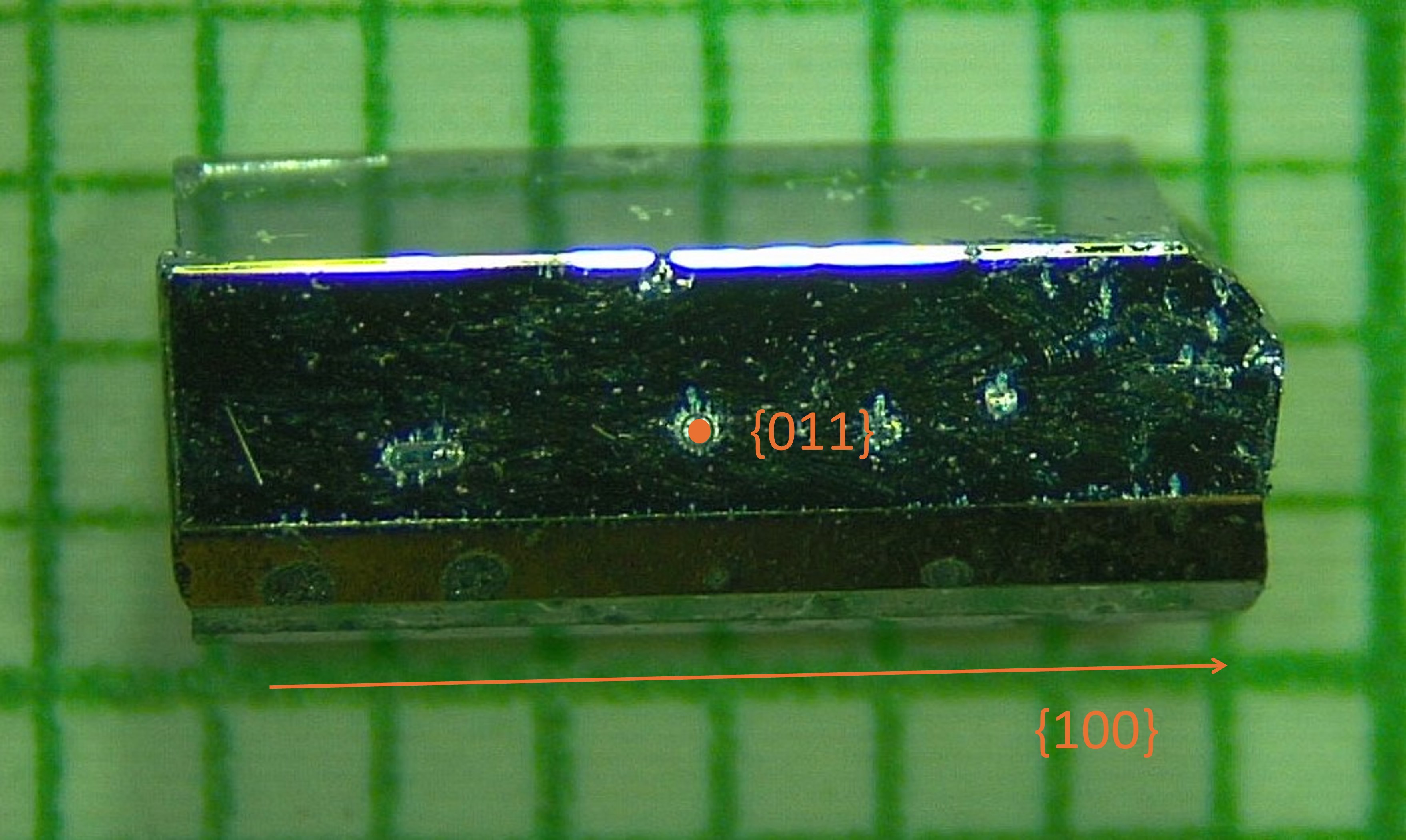}

\caption{\footnotesize{Single crystal of Ho$_2$Co$_6$Al$_{20-\delta}$}}

\protect\label{fig:HoBasedSingleCrystal}
\end{figure}

\begin{figure}
\centering \includegraphics[scale=0.35]{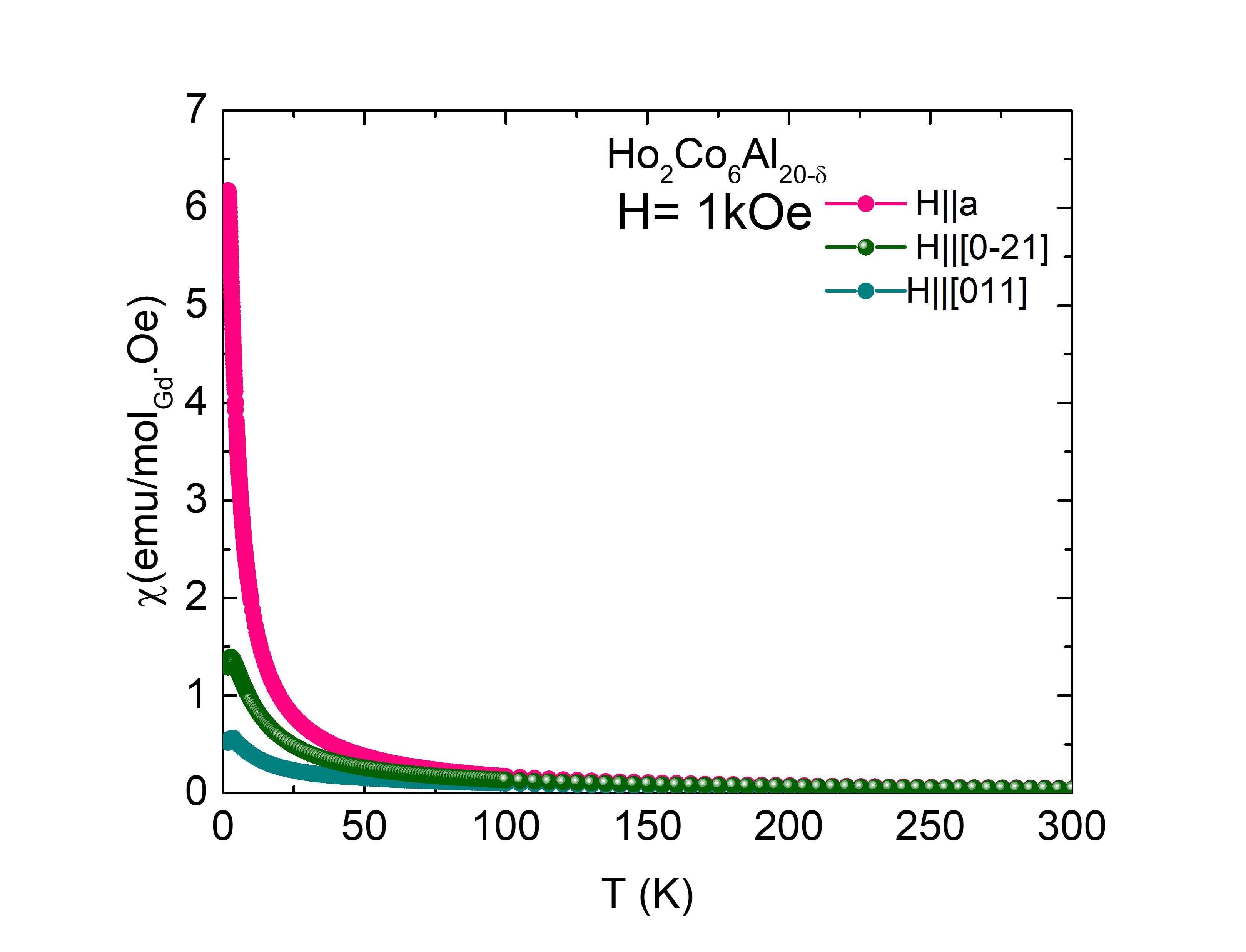}

\caption{\footnotesize{The magnetic temperature susceptibility is plotted as a function of temperature for three different crystallographic directions of Ho$_2$Co$_6$Al$_{20-\delta}$, where [0-21] and [011] directions are perpendicular to [100] direction.} }

\protect\label{fig:HoCoAl_3Directions}
\end{figure}

\begin{figure}
\centering \includegraphics[scale=0.35]{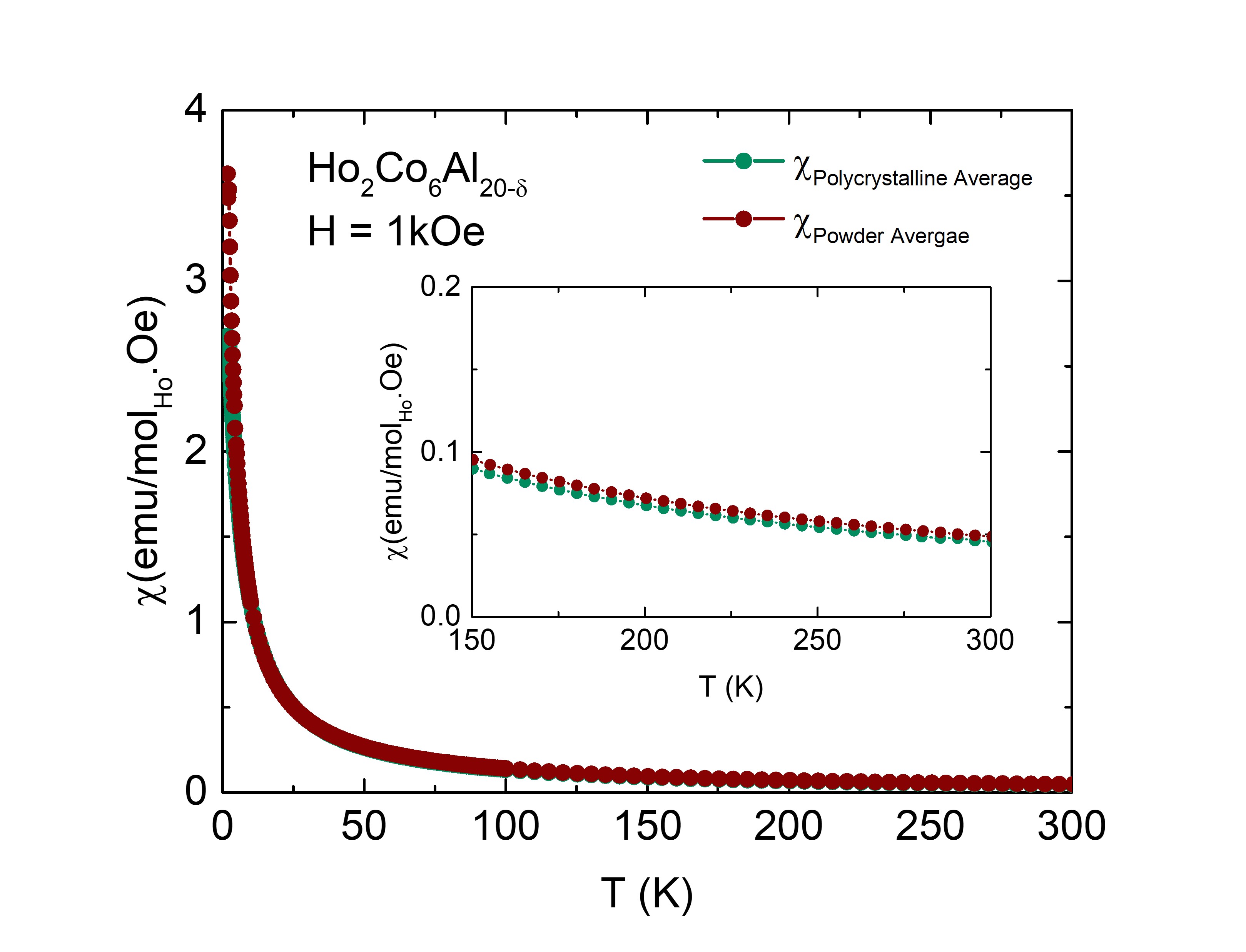}

\caption{\footnotesize{Temperature dependent magnetic susceptibilities \(\chi_{\text{powder}}\) and \(\chi_{\text{poly}}\) for Ho$_2$Co$_6$Al$_{20-\delta}$} were measured at 1 kOe field. The inset highlights the detailed behavior in the high-temperature region.}

\protect\label{fig:HoCoAl_PolyPowder}
\end{figure}

The large single crystal obtained was Ho$_{2}$Co$_{6}$Al$_{20-\delta}$ shown in figure \ref{fig:HoBasedSingleCrystal}. This single crystal was used to perform measurements along three mutually orthogonal directions, as illustrated in figure \ref{fig:HoCoAl_3Directions}. The crystallographic axis was determined by Laue camera, where [0-21] and [011] directions are perpendicular to [100] direction. However, in-plane anisotropy could not be measured for the rest of the compounds because size of our crystals is significantly small and did not allow for reliable in-plane measurements. 
In addition, the direction perpendicular to the \(a\)-axis does not serve as a unique reference for all samples, as achieving consistent alignment along a single axis is experimentally challenging. In any case, the in-plane anisotropy is significantly smaller than the out-of-plane anisotropy. Therefore, it remains feasible to measure the out-of-plane anisotropy.

In figure \ref{fig:HoCoAl_PolyPowder}, we compare the magnetic measurements of the powdered and polycrystalline average obtained from single crystal data. At high temperatures, the two data sets match well as show in the inset, suggesting that the polycrystalline average can serve as a reasonable qualitative approximation to obtain  effective magnetic moment (\(\mu_{\text{eff}}\)). However, at low temperature, noticeable discrepancies emerge between the powdered and the polycrystalline average data. This deviation at low temperature might be because of preferential orientation effects, where the magnetic moments align preferentially along the easy axis, breaking the isotropic assumption that is adopted to calculate the polycrystalline average.   

Despite these challenges, the acquired measurements remain valuable for understanding the anisotropic magnetic properties of this systems, particularly for out-of-plane anisotropy. 
\label{appendixA}

\FloatBarrier

\section*{Appendix B}

\begin{figure}
\centering{}\includegraphics[width=0.4\paperwidth]{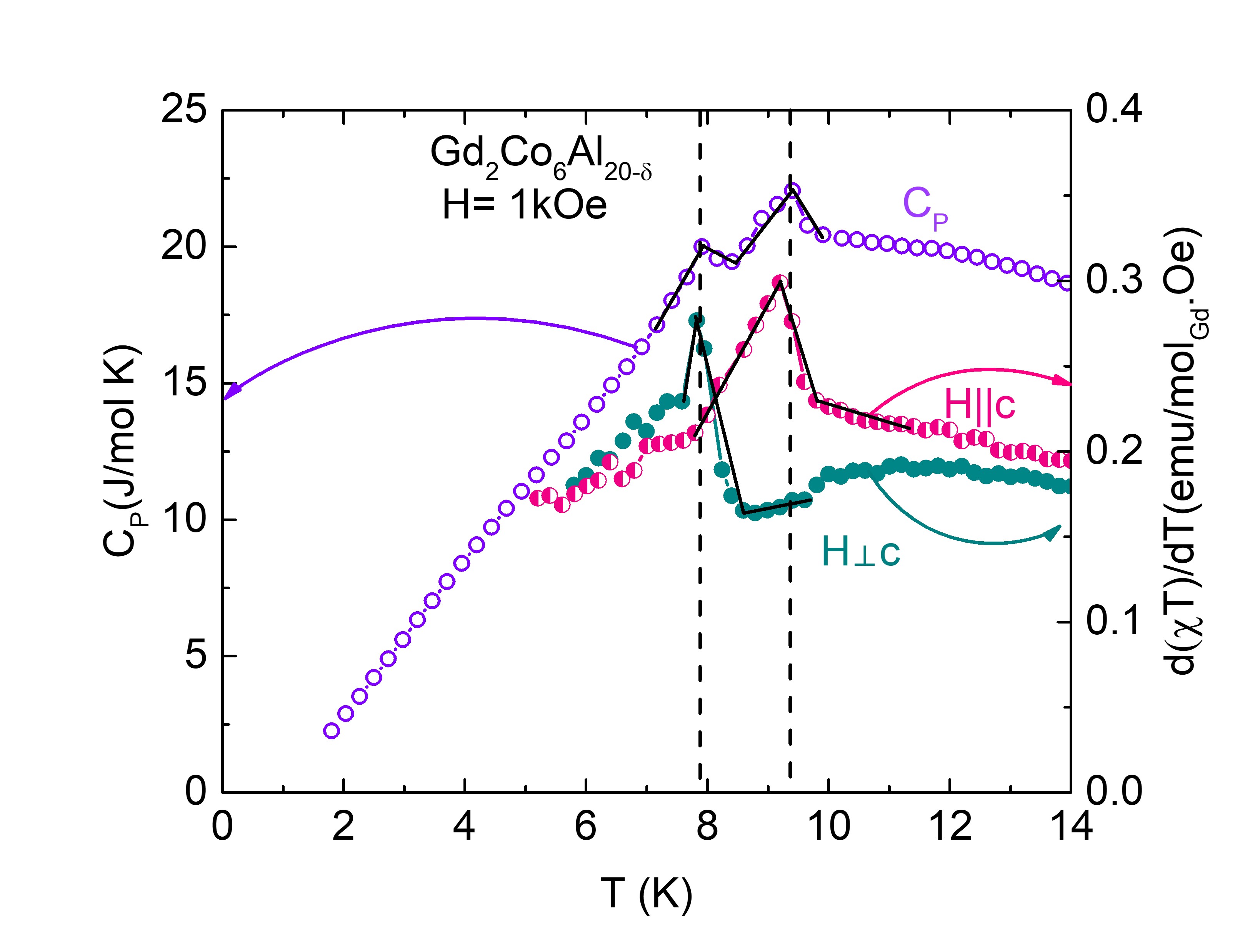}
\caption {\footnotesize{$C_{p}$ (left axis, open symbols) and $d(\chi T)/dT$ (right-axis) obtained for $H\parallel a$ ( half open symbols ) and $H\perp a$ ( solid symbols) with an applied field of $H=1$ kOe. Black solid lines are used to determine transition temperature. \protect\label{fig: transition}}}
\end{figure} 

The criteria used to calculate Néel temperature ($T_N$)  is shown in figure\ref{fig: transition}. The $T_N$ is determined by averaging the onset ($T_{onset}$) and offset ($T_{offset}$) temperatures of the observed peak in the experimental data, such as magnetic susceptibility or specific heat. $T_{onset}$is defined as the temperature at which a deviation from the baseline begins, indicating the start of the magnetic transition. $T_{offset}$ is the maxima of the transition. The Néel temperature is then calculated as:

  \begin{equation}\label{eqn:curie}
T_N = \frac{T_{\text{onset}} + T_{\text{offset}}}{2}
 \end{equation}

The uncertainty in $T_N$ is estimated as half the difference between the onset and offset temperatures:
  \begin{equation}\label{eqn:curie}
 \Delta T_N = \frac{T_{\text{onset}} - T_{\text{offset}}}{2} 
 \end{equation}
\label{appendixB}

\section*{Appendix C}

\begin{figure}
\centering{}\includegraphics[width=0.4\paperwidth]{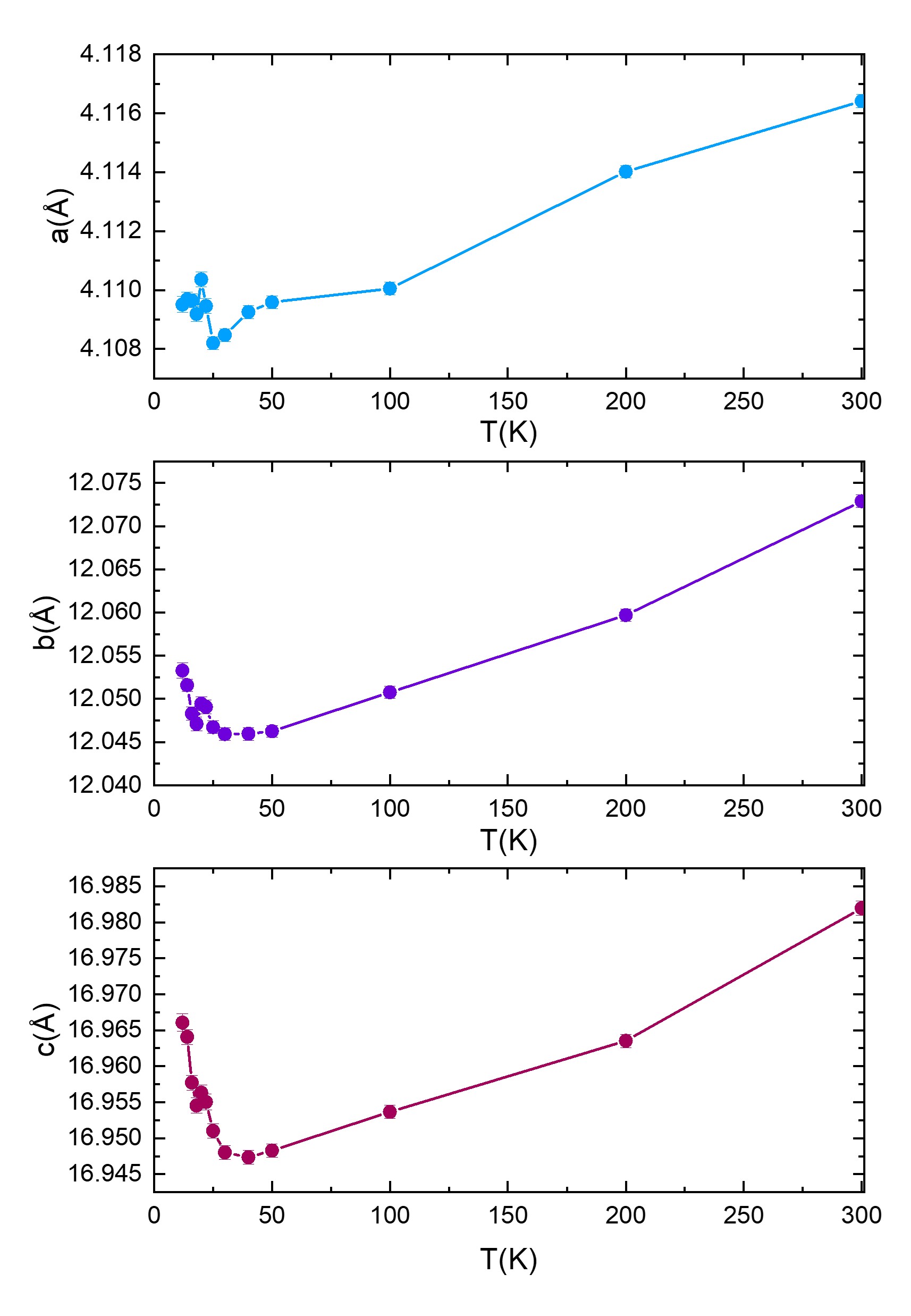}
\caption {\footnotesize{ Evolution of the lattice parameters as a function of temperature, extracted via Rietveld refinement of the temperature-dependent powder X-ray diffraction data of Gd$_2$Co$_6$Al$_{20-\delta}$.
\protect\label{fig: Lattice}}}
\end{figure}

Temperature dependent PXRD measurements were performed from 12 to 300 K using a Rigaku SmartLab X-ray diffractometer equipped with an Oxford K450 helium cryosystem and Cu K$\alpha$ radiation for Gd$_2$Co$_6$Al$_{20-\delta}$. Rietveld refinement was performed for each diffraction pattern using the GSASII software package \citep{von2014small}.The PXRD spectra were first refined at 300 K, and subsequently only the lattice parameters, scale factor, sample displacement, and atomic positions were refined at the remaining temperatures, while keeping all other parameters fixed. 
Upon cooling below $25$ K, clear upturns in the b- and c-lattice parameters (Fig. \ref{fig: Lattice}) and unit cell volume (Fig. \ref{fig:Volume_T}) are seen.  These temperature dependencies are inconsistent with any trivial thermal contraction and instead, when taken with the $\approx$ 18 K $C_p$(T) feature, suggest that there is some form of structural phase transition.  As of yet we do not known what the specific change in basis may be, and further single crystal x-ray and/or neutron diffraction would be needed to resolve this.

\begin{table*}
\centering
\tiny
\renewcommand{\arraystretch}{3}

\caption{Single crystal data and structural refinement information for ${R_2}${Co$_6$}{Al$_{20-\delta}$} ($R$=Gd-Tm) }
\label{tab:CrystalSturcture_Information} %
\begin{tabular}{c|c|c|c|c|c|c|c}
\hline 
\makecell{Chemical \\ formula} & $\text{Gd}_{2}\text{Co}_{6}\text{Al}_{20-\delta}$ & $\text{Tb}_{2}\text{Co}_{6}\text{Al}_{20-\delta}$ & $\text{Y}_{2}\text{Co}_{6}\text{Al}_{20-\delta}$ & $\text{Dy}_{2}\text{Co}_{6}\text{Al}_{20-\delta}$ & $\text{Ho}_{2}\text{Co}_{6}\text{Al}_{20-\delta}$ & $\text{Er}_{2}\text{Co}_{6}\text{Al}_{20-\delta}$ & $\text{Tm}_{2}\text{Co}_{6}\text{Al}_{20-\delta}$ \\
\hline 
\hline
 $\delta$ & 0.91  & 0.86 & 0.86 & 0.73  & 0.83   & 0.79  & 0.77 \\
\hline 
 Formula weight(g/mol)  & 1183.26  & 1187.68 & 1040.37 & 1197.27  & 1200.65   & 1205.85  & 1210.27 \\
\hline 
Temperature  & 298(1)  & 294.53(18) & 295.43(19) & 295.15(4) & 298.3(9) & 294.2(6) & 295.8(9)  \\
\hline 
 Wavelength(\AA,Ag K$\alpha$)  & 0.56087  & 0.56087 & 0.56087 & 0.56087 & 0.56087 & 0.56087& 0.56087 \\
\hline 
Crystal system  & Orthorhombic & Orthorhombic & Orthorhombic& Orthorhombic & Orthorhombic & Orthorhombic & Orthorhombic \\
\hline 
Space group  & Imma & Imma & Imma & Imma& Imma & Imma & Imma\\
\hline 
Unit cell dimensions  & \textit{a} = 4.1215(2) \AA & \textit{a} = 4.1097(1) \AA & \textit{a} = 4.11080(10) \AA & \textit{a} = 4.1089(1) \AA & \textit{a} = 4.1015(2) \AA & \textit{a} = 4.0981(2) \AA & \textit{a} = 4.0976(1) \AA \\
        & \textit{b} = 12.0831(5) \AA & \textit{b} = 12.0657(4) \AA & \textit{b} = 12.0710(3) \AA &\textit{b} = 12.0572(3) \AA & \textit{b} = 12.0520(5) \AA & \textit{b} = 12.0249(6) \AA & \textit{b} = 12.0451(4) \AA \\
        & \textit{c} = 16.9888(6) \AA & \textit{c} = 16.9702(4) \AA & \textit{c} = 16.9554(5) & \textit{c} = 16.9486(4) \AA & \textit{c} = 16.9283(8) \AA & \textit{c} = 16.9372(9) \AA & \textit{c} = 16.9069(5) \AA \\
                  & $\alpha$ = 90° & $\alpha$ = 90°& $\alpha$ = 90°& $\alpha$ = 90°& $\alpha$ = 90°& $\alpha$ = 90°& $\alpha$ = 90°\\
                   & $\beta$ = 90° & $\beta$ = 90°& $\beta$ = 90°& $\beta$ = 90°& $\beta$ = 90°& $\beta$ = 90°& $\beta$ = 90°\\
                    & $\gamma$ = 90° & $\gamma$ = 90°& $\gamma$ = 90°& $\gamma$ = 90°& $\gamma$ = 90°& $\gamma$ = 90°& $\gamma$ = 90°\\

\hline 
Volume($cm^3$)  & 846.05(6) & 841.49(4)& 841.35(4) & 839.66(4) & 836.79(7)& 834.65(7) & 834.46(4) \\
\hline 
Z & 2 & 2  & 2 & 2 & 2  & 2  & 2 \\
\hline 
Calculated density($g/cm^3$)  & 4.645 & 4.687 & 4.107 & 4.735 & 4.765 & 4.798 & 4.817 \\
\hline 
Absorption coefficient (\(\text{mm}^{-1}\)) & 7.531 & 7.839 & 7.004 & 8.131 & 8.444 & 8.767 & 9.083 \\
\hline 
Absorption Correction  & multi-scan  & multi-scan  & multi-scan & multi-scan & multi-scan & multi-scan  & multi-scan \\
\hline 
 F(000)  & 1076.0  & 1082.0 & 971.0 & 1088.0  & 1090.0   & 1095.0 & 1100.0 \\
\hline 
Crystal size(\text{mm}) & 0.07 $\times$ 0.05 $\times$0.04  & 0.11 $\times$ 0.09 $\times$0.07 & 0.18 $\times$ 0.14 $\times$0.07  & 0.1 $\times$ 0.08 $\times$0.06  & 0.04 $\times$ 0.03 $\times$0.03 & 0.11 $\times$ 0.06 $\times$0.05 & 0.11 $\times$ 0.08 $\times$0.07 \\
\hline 
theta range for data collection  & 2.66$\degree$ - 31.68$\degree$ & 2.664$\degree$ - 31.812$\degree$  & 2.663$\degree$ - 31.529$\degree$  & 2.666$\degree$ - 31.732$\degree$  & 2.667$\degree$ - 31.996$\degree$   & 2.673$\degree$ - 31.994$\degree$ & 2.669$\degree$-31.832$\degree$ \\
\hline 
Index ranges (min/max,h,k,l) & [-7/5, -21/16, -30/28] & [-7/7, -20/20, -31/25] & [-7/6, -20/22, -30/30] & [-7/7, -21/21, -31/28]  &  [-7/6,-22/21,-30/30]  & [-7/6, -22/21, -30/29] & [-7/7, -22/21, -31/30] \\
\hline 
Reflections collected  & 5172  & 10883 & 11231 & 11405 &  11364 & 11275 & 10243 \\
\hline 
Independent reflections & 1354[$R_{int} = 0.0244$] & 1461[$R_{int} = 0.0177$] & 1498[$R_{int} = 0.0294$] & 1490[$R_{int} = 0.0188$]  & 1498[$R_{int} = 0.0333$] & 1489[$R_{int} = 0.0232$] & 1505[$R_{int} = 0.0185$] \\ 
\hline 
Data /restrains /parameters  & 1354/4/52  & 1461/4/52 & 1498/4/54 & 1490/4/52 &  1498/4/52 & 1489/4/52 & 1505/4/52 \\
\hline 
GOF & 1.068  & 1.135 & 0.958 &1.088 & 1.064 & 1.172 & 1.132 \\
\hline 
Final R indices[I$>$2$\sigma$(I)]  & {$R_1 = 0.0269$,} & {$R_1 = 0.0182$,} & {$R_1 = 0.0229$,} & {$R_1 = 0.0199$,} & $R_1 = 0.0206$,&  $R_1 = 0.0182$, & {$R_1 = 0.0195$,} \\
& $wR_2 = 0.0657$ & $wR_2 = 0.0494$ & $wR_2 = 0.0538$ & $wR_2 = 0.0494$ & $wR_2 = 0.0458$  & $wR_2 = 0.0486$ & $wR_2 = 0.0489$\\
\hline 
 R indices[all data] & {$R_1 = 0.0311$},   & {$R_1 = 0.0191$}, & {$R_1 = 0.0258$ },& {$R_1 = 0.0208$ }, & {$R_1 = 0.0232$},   &  $R_1 = 0.0196$, & {$R_1 = 0.0210$}, \\
 & $wR_2 = 0.0675$ & $wR_2 = 0.0479$ & $wR_2 = 0.0548$ & $wR_2 = 0.0497$ & $wR_2 = 0.0466$ & $wR_2 = 0.0490$ & $wR_2 = 0.0494$\\
 
\hline 
Largest diff. peak  & 3.42/-1.99  & 2.10/-1.73 & 2.63/-1.19 & 2.75/-1.97 & 2.12/-1.39 & 2.03/-1.71 & 2.36/-1.75 \\ 
and hole ($e^{-}/\text{\AA}^3$) & & & & & &\\
\hline

\hline
\end{tabular}
\end{table*}

In tables from \ref{tab:scxrd_Tb}-\ref{tab:scxrd_Tm}, a comprehensive set of crystallographic parameters derived from SCXRD refinements is presented. Atomic positions, occupancies and thermal displacement parameters for the $R$$_{2}$Co$_{6}$Al$_{20-\delta}$.
\begin{table*}
\centering
\caption{ Refined atomic positions, site occupancy, and isotropic thermal displacement parameters for Tb$_{2}$Co$_6$Al$_{20-\delta}$, $\delta\sim 1$, according to SCXRD.}

\begin{tabular}{lcccccc}

At. & Site & $x$ & $y$ & $z$ & Occ. & $U_{\text
{iso}}$\\
\hline
Tb & $4e$ & 1 & 3/4 & 0.46982(2) & 1 & 0.00694(4) \\
Co & $4e$ & 0 & 1/4 & 0.81258(3) & 1 & 0.00404(7)\\
Co & $8h$ & 0 & 0.55245(3) & 0.63200(2) & 1 & 0.00575(6) \\
Al & $4b$ & 0 & 0.36698(7) & 0.69603(4) & 1 & 0.00561(11) \\
Al & $8h$ & 0 & 1/2 & 1/2 & 1 & 0.00944(18) \\
Al & $8h$ & 0 & 0.42003(7) & 0.89467(5) & 1 & 0.00923(13)\\
Al & $8h$ & 1/2 & 0.63584(7) & 0.57451(6) & 1 & 0.01170(14) \\
Al & $4e$ & 0 & 3/4 & 0.66501(8) & 1 & 0.0118(2) \\
Al & $8h$ & 0 & 0.5773(2) & 0.77869(13) & 0.455(4) & 0.0253(5) \\
Al & $16j$ & 0.1642(19) & 0.5792(4) & 0.7620(2) & 0.164(3) & 0.0253(10) \\
\label{tab:scxrd_Tb} 
\end{tabular}
\end{table*}
\medskip

\begin{table*}
\centering
\caption{ Refined atomic positions, site occupancy, and isotropic thermal displacement parameters for Dy$_{2}$Co$_6$Al$_{20-\delta}$, $\delta\sim 1$, according to SCXRD.}

\begin{tabular}{lcccccc}

At. & Site & $x$ & $y$ & $z$ & Occ. & $U_{\text
{iso}}$\\
\hline
Dy & $4e$ & 1 & 3/4 & 0.46966(2) & 1 & 0.00732(4) \\
Co & $4e$ & 0 & 1/4 & 0.81269(3) & 1 & 0.00398(7)\\
Co & $8h$ & 0 & 0.55261(3) & 0.63204(2) & 1 & 0.00567(6) \\
Al & $4b$ & 0 & 0.36694(7) & 0.69594(5) & 1 & 0.00561(13) \\
Al & $8h$ & 0 & 1/2 & 1/2 & 1 & 0.0098(2) \\
Al & $8h$ & 0 & 0.42011(8) & 0.89495(6) & 1 & 0.00954(15)\\
Al & $8h$ & 1/2 & 0.63596(8) & 0.57399(7) & 1 & 0.01100(15) \\
Al & $4e$ & 0 & 3/4 & 0.66486(8) & 1 & 0.0115(2) \\
Al & $8h$ & 0 & 0.5779(2) & 0.77907(14) & 0.463(4) & 0.0244(6) \\
Al & $16j$ & 0.1593(19) & 0.5789(4) & 0.7628(2) & 0.172(3) & 0.0244(11) \\
\label{tab:scxrd_Dy} 
\end{tabular}
\end{table*}
\medskip

\begin{table*}
\centering
\caption{ Refined atomic positions, site occupancy, and isotropic thermal displacement parameters for Ho$_{2}$Co$_6$Al$_{20-\delta}$, $\delta\sim 1$, according to SCXRD.}

\begin{tabular}{lcccccc}
At. & Site & $x$ & $y$ & $z$ & Occ. & $U_{\text
{iso}}$\\
\hline
Ho & $4e$ & 1 & 3/4 & 0.46961(2) & 1 & 0.00758(4) \\
Co & $4e$ & 0 & 1/4 & 0.81281(3) & 1 & 0.00413(7) \\
Co & $8h$ & 0 & 0.55276(3 & 0.63204(2) & 1 & 0.00600(6) \\
Al & $4b$ & 0 & 0.36695(7) & 0.69586(5) & 1 & 0.00592(12) \\
Al & $8h$ & 0 & 1/2 & 1/2 & 1 & 0.0095(2) \\
Al & $8h$ & 0 & 0.41983(7) & 0.89525(6) & 1 & 0.00949(14) \\
Al & $8h$ & 1/2 & 0.63608(7) & 0.57376(6) & 1 & 0.01105(15) \\
Al & $4e$ & 0 & 3/4 & 0.66486(8) & 1 & 0.0115(2) \\
Al & $8h$ & 0 & 0.5775(2) & 0.77918(13) & 0.455(4) & 0.0236(5) \\
Al & $16j$ & 0.1589(17) & 0.5792(4) & 0.7628(2) & 0.169(3) & 0.0236(10) \\
\label{tab:scxrd_Ho} 
\end{tabular}
\end{table*}
\medskip

\begin{table*}
\centering
\caption{ Refined atomic positions, site occupancy, and isotropic thermal displacement parameters for Er$_{2}$Co$_6$Al$_{20-\delta}$, $\delta\sim 1$, according to SCXRD.}

\begin{tabular}{lcccccc}
At. & Site & $x$ & $y$ & $z$ & Occ. & $U_{\text
{iso}}$\\
\hline
Er & $4e$ & 1 & 3/4 & 0.46946(2) & 1 & 0.00728(4) \\
Co & $4e$ & 0 & 1/4 & 0.81292(3) & 1 & 0.00400(7) \\
Co & $8h$ & 0 & 0.55290(3) & 0.63208(2) & 1 & 0.00587(6) \\
Al & $4b$ & 0 & 0.36697(7) & 0.69579(5) & 1 & 0.00555(12) \\
Al & $8h$ & 0 & 1/2 & 1/2 & 1 & 0.00897(19) \\
Al & $8h$ & 0 & 0.41964(8) & 0.89553(6) & 1 & 0.00906(14) \\
Al & $8h$ & 1/2 & 0.63607(7) & 0.57331(6) & 1 & 0.01064(15) \\
Al & $4e$ & 0 & 3/4 & 0.66482(8) & 1 & 0.0110(2) \\
Al & $8h$ & 0 & 0.5778(2) & 0.77948(13) & 0.461(4) & 0.0234(5) \\
Al & $16j$ & 0.1554(18) & 0.5786(4) & 0.7634(2) & 0.171(3) & 0.0234(10) \\

\label{tab:scxrd_Er} 
\end{tabular}
\end{table*}
\medskip

\begin{table*}
\centering
\caption{ Refined atomic positions, site occupancy, and isotropic thermal displacement parameters for Tm$_{2}$Co$_6$Al$_{20-\delta}$, $\delta\sim 1$, according to SCXRD.}

\begin{tabular}{lcccccc}

At. & Site & $x$ & $y$ & $z$ & Occ. & $U_{\text
{iso}}$\\
\hline
Tm & $4e$ & 1 & 3/4 & 0.46920(2) & 1 & 0.00776(4) \\
Co & $4e$ & 0 & 1/4 & 0.81304(3) & 1 & 0.00418(7) \\
Co & $8h$ & 0 & 0.55298(3) & 0.63208(2) & 1 & 0.00604(6) \\
Al & $4b$ & 0 & 0.36697(7) & 0.69573(5) & 1 & 0.00581(13) \\
Al & $8h$ & 0 & 1/2 & 1/2 & 1 & 0.0094(2) \\
Al & $8h$ & 0 & 0.41971(8) & 0.89564(6) & 1 & 0.00942(14) \\
Al & $8h$ & 1/2 & 0.63618(8) & 0.57296(6) & 1 & 0.01069(15) \\
Al & $4e$ & 0 & 3/4 & 0.66464(8) & 1 & 0.0111(2) \\
Al & $8h$ & 0 & 0.5780(2) & 0.77970(14) & 0.460(4) & 0.0227(5) \\
Al & $16j$ & 0.1508(18) & 0.5788(4) & 0.7640(2) & 0.174(3) & 0.0227(10) \\
\label{tab:scxrd_Tm} 
\end{tabular}
\end{table*}

\begin{table*}
\centering
\caption{ Refined atomic positions, site occupancy, and isotropic thermal displacement parameters for Y$_{2}$Co$_6$Al$_{20-\delta}$, $\delta\sim 1$, according to SCXRD.}
\begin{tabular}{lcccccc}
At. & Site & $x$ & $y$ & $z$ & Occ. & $U_{\text
{iso}}$\\
\hline
Y & $4e$ & 1 & 3/4 & 0.46972(2) & 1 & 0.00763(7) \\
Co & $4e$ & 0 & 1/4 & 0.81268(2) & 1 & 0.00526(7) \\
Co & $8h$ & 0 & 0.55257(2) & 0.63204(2) & 1 & 0.00708(7) \\
Al & $4b$ & 0 & 0.36685(6) & 0.69609(4) & 1 & 0.00712(11) \\
Al & $8h$ & 0 & 1/2 & 1/2 & 1 & 0.01187(17) \\
Al & $8h$ & 0 & 0.42059(6) & 0.89438(5) & 1 & 0.01124(13) \\
Al & $8h$ & 1/2 & 0.63593(6) & 0.57451(5) & 1 & 0.01291(14) \\
Al & $4e$ & 0 & 3/4 & 0.66522(7) & 1 & 0.01293(18) \\
Al & $8h$ & 0 & 0.57746(18) & 0.77863(11) & 0.454(3) & 0.0262(4) \\
Al & $16j$ & 0.1625(17) & 0.5793(3) & 0.76169(19) & 0.165(3) & 0.0262(8) \\
\label{tab:scxrd_Tm} 
\end{tabular}
\end{table*}
\FloatBarrier

\section*{Appendix D}
The specific heat of the isostructural, non-magnetic compound $\text{Y}_2\text{Co}_6\text{Al}_{20-\delta}$ was used as a reference to estimate the combined electronic and lattice contributions to the total specific heat ($C_p$) of the $R_2\text{Co}_6\text{Al}_{20-\delta}$ series. This procedure allows for the isolation of the magnetic contribution, $C_{\text{mag}}$, and the subsequent evaluation of the magnetic entropy, $S_{\text{mag}}$. To account for the atomic mass mismatch between Y and the rare-earth ions, a mass-correction scaling of the Debye temperature ($\Theta_D$) was evaluated. The Debye temperature for the Y-reference, $\Theta_D^Y$, was determined from low-temperature specific heat data by fitting the electronic ($\gamma$) and phonon ($\beta$) coefficients using the relation $C_p/T = \gamma + \beta T^2$.The Debye temperature was calculated using equation \ref{eqn:theta}.

\begin{equation} \label{eqn:theta}\Theta_D = \left( \frac{12\pi^4 nR}{5\beta} \right)^{1/3} ,\end{equation}
where $n$ represents the number of atoms per formula unit. To estimate the lattice contribution for each $R$ member, $\Theta_D^R$ was scaled according to equation \ref{eqn:Masscorrection}.

\begin{equation} \label{eqn:Masscorrection}\Theta_D^R = \Theta_D^Y \left( \frac{M_Y}{M_R} \right)^{1/2}\end{equation}

where $M_Y$ and $M_R$ are the respective molar masses. While this scaling was used to determine a re-normalized phonon coefficient $\beta$ that was used to recalculate $C_P^Y$ and used in equation \ref{eqn: Cp} to calculate $C_{mag}$. The resulting corrections to $C_{\text{mag}}$ were found to be negligible. Consequently, the magnetic specific heat and entropy were determined using equation \ref{eqn: Cp} and \ref{eqn: S}.

\begin{equation}\label{eqn: Cp}
C_{\text{mag}} = C_p^R - C_p^Y\end{equation}

\begin{equation} \label{eqn: S}
S_{\text{mag}}(T) = \int_{0}^{T} \frac{C_{\text{mag}}}{T'} dT'.\end{equation}

For $\text{Er}_2\text{Co}_6\text{Al}_{20-\delta}$, the calculated $S_{\text{mag}}$ values carry a higher degree of uncertainty. The absence of a resolved magnetic transition within the experimental temperature range, combined with a low-temperature divergence (upturn) in $C_p/T$, necessitated a manual extrapolation to $T = 0$ to estimate the total entropy.

\bibliography{2024R2Co6Al19_references}

\end{document}